\documentclass[aps,pra,reprint,twocolumn,showpacs,showkeys,floatfix,nobibnotes]{revtex4-1}

\usepackage{hyperref}
\hypersetup{colorlinks=true, urlcolor=red, citecolor=blue, linkcolor=blue}
\usepackage{graphicx,epstopdf}
\usepackage{subfigure}
\usepackage{epstopdf}
\usepackage{lastpage}
\usepackage{braket}

\begin{document}
\title{Lower- versus Higher-order nonclassicalities for a coherent superposed quantum state}
\author{Deepak}
\author{Arpita Chatterjee}
\email{Corresponding author: arpita.sps@gmail.com}

\affiliation{Department of Mathematics, J. C. Bose University of Science and Technology, YMCA, Faridabad 121006, Haryana, India}

\date{\today}

\begin{abstract}

A coherent state is defined conventionally in different ways such as a displaced vacuum state, an eigenket of annihilation operator or as an infinite dimensional Poissonian superposition of Fock states. In this work, we describe a superposition $(ta+ra^\dagger)$ of field annihilation and creation operators acting on a continuous variable coherent state $\ket{\alpha}$ and specify it by $\ket\psi$. We analyze the lower- as well as the higher-order nonclassical properties of $\ket\psi$. The comparison is performed by using a set of nonclassicality witnesses (e.g., higher-order photon-statistics, higher-order antibunching, higher-order sub-Poissonian statistics, higher-order squeezing, Agarwal-Tara parameter, Klyshko's condition and a relatively new concept, matrix of phase-space distribution). It is found that higher-order criteria are much more efficient to detect the presence of nonclassicality as compared to lower-order conditions.

\end{abstract}



\maketitle

\section{Introduction}

Coherent state, a specific quantum state introduced by Glauber \cite{glauber} using the harmonic oscillator algebra, has been a leading field of interest in the quantum optics and atom optics community for a number of reasons. For example, a coherent state can be used to solve the quantum mechanical problem of a harmonic oscillator acted on by a time-dependent force. In the quantum theory, a coherent state can be employed to describe a wide range of physical systems like the oscillating motion of a particle confined in a quadratic potential well, a state in a system for which the ground-state wave-packet is displaced from the origin of the system etc. Another application of coherent state can be found in the context of the sensitivity limit imposed by the quantum mechanics on detectors for gravitational radiation \cite{agar,klauder}. There are many other fields of applications of canonical coherent states, ranging from quantization to signal processing and image processing. In chemistry, linear superposition of coherent states is used in order to construct multidimensional wavefunctions. In the field of biology, these can be used to describe the long-range forces between human blood cells and the long-range phase coherence in the bacteriorhodospin macromolecules \cite{zhang}. With the advent of quantum state engineering \cite{bellini,xu,chat}, quantum computing and communication (\cite{priya1,pathak2} and references therein), a large number of theoretical as well as experimental strategies have been proposed for manufacturing and controlling various types of coherent states \cite{tiwari,dajka}.

Manipulation of a light field at the single-photon level provides a promising area for many important applications in quantum information science \cite{tittel,knill}, such as  non-Gaussian two-mode entangled states are used for a nonlocality test \cite{carmichael} and entanglement distillation \cite{taka}, photon-added squeezed states are suggested to improve the fidelity of continuous variable (CV) teleportation \cite{yang} etc. In particular, two elementary operations on a single-mode field (i.e., photon subtraction
and addition represented by bosonic annihilation and creation operators $a$ and $a^\dagger$, respectively) can be employed to transform
a field state to a desired one \cite{kim}. For example, Agarwal and Tara \cite{tara1} proposed theoretically a non-Gaussian, nonclassical state, which is intermediate between the coherent state $\ket\alpha$ (most classical-like quantum state) and the number state $\ket n$ (purely quantum state), by repeated application of the photon creation operator on the coherent state basis. The nonlinear coherent state or $f$-coherent state ${\ket f}_\alpha$ was introduced by \cite{filho,manko} as eigenstate of a deformed annihilation operator $A{\ket\alpha}_f = \alpha{\ket\alpha}_f$ where $A=a f(N)$, $f(N)$ being a deformation function of the number operator $N={a}^\dagger a$, and also by the application of a deformed displacement operator upon the vacuum state, such as ${\ket\alpha}_D=D_D(\alpha)\ket 0$ \cite{ghosh1}. Another idea was developed by Kim et. al. \cite{parigi} to implement a coherent superposition $a a^\dagger+a^\dagger a$ of two-product operations $a a^\dagger$ and $a^\dagger a$. Later Lee and Nha considered a coherent superposition of photonic operations at a more elementary level; that is, the superposition of photon subtraction and addition, $ta+ra^\dagger$, and
investigated how it transforms a classical coherent state to a nonclassical one \cite{lee}. Furthermore, they introduced an interference set-up to realize this coherent operation in an optical experiment and employed it together with displacement operators to generate an arbitrary superposition of number states involving up to two photons. The superposition state $c_0\ket 1+c_1 \ket 1+c_2\ket 2$ can be used for quantum information processing; for example, the nonlinear sign-shift (NS) gate (a basic element of the CNOT
gate) \cite{ralph}, and the optimal estimation of the loss parameter of a bosonic channel \cite{adesso}. We extend the concept of Lee and Nha by studying the higher-order nonclassical properties of a state generated by applying $ta+ra^\dagger$ over input $\ket\alpha$.

A quantum state is defined as nonclassical (i.e. a state having no classical analogue) if its Glauber-Sudarshan $P$-function has negative values. Unfortunately, except a single proposal for the measurement of $P$-function in a special case \cite{kiesel}, there is no method for experimental determination of $P$-function. Thus a number of feasible criteria for witnessing nonclassicality has been developed (\cite{miran,monica} and references therein). These nonclassicality witnesses can be expressed in terms of moments of annihilation and creation operators. If the moments include terms up to fourth orders of $a$ and $a^\dagger$ (i.e. second-order correlations), the corresponding nonclassical feature is referred as lower-order nonclassicality. As a consequence, higher-order nonclassicality is related with the conditions observed via higher-order correlations. Most frequently studied higher-order nonclassical features are higher-order antibunching (HOA) \cite{gracia}, higher-order sub-Poissonian photon
statistics (HOSPS) \cite{prakash}, higher-order squeezing (HOS) of Hillery type \cite{hillery} and Hong-Mandel type \cite{hong1} etc. The experimental success in detecting higher-order nonclassicality and the fact that weaker nonclassicality not detected by lower-order criteria can be spotted by their higher-order counterparts have led to a large number of theoretical works in this direction \cite{allevi1,jack}. In fact, HOA has been reported in optomechanical and optomechanical-like system
\cite{alam2}, optical coupler \cite{kishore1}, hyper Raman process \cite{kishore4} etc., HOSPS has been reported in finite dimensional coherent
state \cite{alam1}, photon added and subtracted squeezed coherent states \cite{kishore2} etc., and HOS has been reported in finite dimensional coherent state \cite{alam1} and a pair of anharmonic oscillators \cite{alam3}. However, the fact that no effort (to the best of our knowledge) has been made so far to investigate the higher-order nonclassical properties of a superposed coherent state $(ta+ra^\dagger)\ket\alpha$, is motivated us to work on it. We have also employed a very recent approach for certifying the nonclassical features of $\ket\psi$ via correlations of phase-space distributions \cite{martin}.

The paper is structured as follows: we describe the general theory for the superposed coherent state $\ket\psi$ in Section~\ref{sec2}. The next section illustrates different higher-order nonclassical criteria of $\ket\psi$ and matrix of phase-space distribution of it. Section~\ref{sec3} ends with a summary of the main results of this article.

\section{General theory for our quantum state of interest}
\label{sec2}

In this section, we focus on a coherent superposition of elementary photonic operations, that means, the superposition of photon subtraction and addition $ta+ra^\dagger$, $t$ and $r$ are scalars with $t^2+r^2=1$. Given a coherent state $\ket\alpha$ as an input field, the superposed state can be described as \cite{lee}
\begin{eqnarray}
\label{eq1}
\ket{\psi} = N^{-1/2}(ta+ra^\dagger)\ket{\alpha},
\end{eqnarray}
where $N = \Big[r^2 +|\alpha|^2 +rt(\alpha^2+\alpha^{*2})\Big]$ is the normalization constant.

\begin{figure*}[ht]
\centering
\includegraphics[width=8cm]{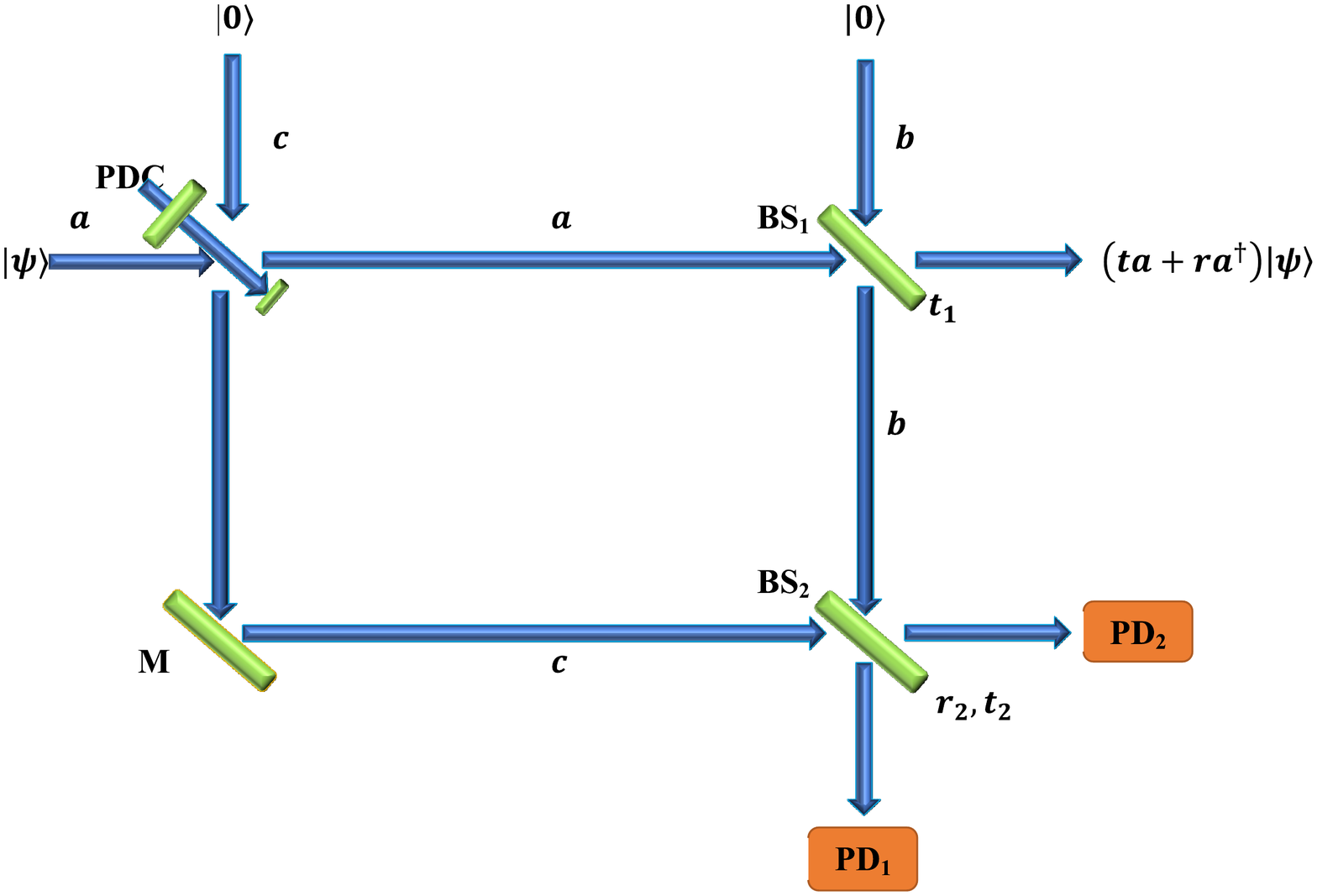}

\caption{(Color online) An illustration for the experimental proposal of $ta+ra^\dagger$.}
\label{fig01}
\end{figure*}

The generation of the desired quantum operation $ta+ra^\dagger$ involves proper sequencing of photon subtraction and photon addition operators, and then coherent superposition of them by removing the which-path information \cite{himadri1}. An experimental scheme for generating this quantum operation is shown in Fig.~\ref{fig01}.

A high-transmissivity beam-splitter $\mathrm{BS}_1$ is used here for photon subtraction. When an arbitrary input field $\ket\psi$ is injected into a high-transmissivity beam-splitter with the other input in a vacuum mode, the detection of a photon in the photodetector implies that a single photon is subtracted from the initial state, due to the conservation of photon number. This corresponds to the action of $a\ket\psi$, which holds well particularly when the transmissivity $t_1$ of $\mathrm{BS}_1$ is large enough \cite{wenger}. A parametric down-converter (PDC) is used to add photon. If the initial state is injected to a signal mode of a PDC with the idler mode in a vacuum state, the detection of a single photon at the output idler mode heralds that one photon is added to the input state, due to the pairwise photon-creation and destruction mechanism of the PDC. This corresponds to the action $a^\dagger\ket\psi$, which holds well particularly when the interaction strength in the PDC is small \cite{zavatta}. An additional beam-splitter $\mathrm{BS}_2$ with transmissivity $t_2$ and reflectivity $r_2$ is used to erase the which-path information on the detected single photon. $\mathrm{M}$ is a highly reflective mirror and $\mathrm{PD}_1$, $\mathrm{PD}_2$ are the photodetectors, which detect the success of the addition or subtraction process in an optical path.

The generation of a coherent superposed state can be described mathematically using standard operators for the various paths involved in the scheme. In Fig.~\ref{fig01}, an arbitrary state $\ket\psi$ is injected into the parametric down converter with small coupling strength $\eta\ll 1$, which acts as
\begin{eqnarray*}
e^{(-\eta a^\dagger c^\dagger + \eta ac)}\ket{\psi}_a\ket{0}_c \approx (1 - \eta a^\dagger c^\dagger)\ket{\psi}_a\ket{0}_c
\end{eqnarray*}
Next, the state is incident upon a beam-splitter $\mathrm{BS}_1$ (transmissivity $t_1\approx 1$). The resulting operation can be written as
\begin{eqnarray*}
{B}_{1ab}(1 - \eta a^\dagger c^\dagger)\ket{\psi}_a\ket{0}_b\ket{0}_c
& \approx & \left(1 - \frac{r^*_1}{t_1}ab^\dagger\right)(1 - \eta a^\dagger c^\dagger)\\
& & \ket{\psi}_a\ket{0}_b\ket{0}_c
\end{eqnarray*}
The second beam-splitter $\mathrm{BS}_2$ with the transformations $b'=t_2 b+r_2 c$ and $c'= t^*_2 c-r^*_2 b$ is used to remove the path information and produce the superposition state. Here $b$ and $c$ ($b'$ and $c'$) are the input (output) modes of the beam-splitter. Using the above relations, $\mathrm{BS}_2$ yields
\begin{eqnarray*}
\begin{array}{lcl}
& & B_{2bc} B_{1ab}(1 - \eta a^\dagger c^\dagger)\ket{\psi}_a\ket{0}_b\ket{0}_c\\\\
& \equiv & \left\{1 - \frac{r^*_1}{t_1}a(t_2^*b^\dagger + r_2^* c^\dagger)\right\}\left\{1 - \eta a^\dagger (t_2 c^\dagger-r_2b^\dagger)\right\}\\
& & \ket{\psi}_a\ket{0}_b\ket{0}_c
\end{array}
\end{eqnarray*}
The detection of single photon at $\mathrm{PD}_1\,\,(\mathrm{PD}_2)$ and no photon at $\mathrm{PD}_2\,\,(\mathrm{PD}_1)$ leads to the state $(t a+r a^\dagger)\ket\psi$ with $t \approx -\frac{r_1^*t_2^*}{t_1}$ $\left(-\frac{r_1^*r_2^*}{t_1}\right)$ and $r \approx -\eta t_2$ ($\eta r_2$).

\section{Nonclassical features of the superposed state}
\label{sec3}

An arbitrary quantum state is named as nonclassical if its Glauber-Sudarshan $P$-function fails to be a classical probability distribution \cite{glauber, sudarshan}. That means the negative value of $P$-function suggests that the state is not enjoying classical status, and can be considered as a nonclassical one. Since there is no direct measurement for $P$-function, many operational criteria, such as, negative values of Wigner function \cite{wigner,kenfack}, zeros of $Q$ function \cite{husimi,stephen}, several moment-based measures \cite{miran, monica} have been proposed for identification of nonclassicality. Most of these conditions are one-sided only in the sense that if a criteria is satisfied then the state is definitely nonclassical but when the condition is not satisfied, one cannot conclude about the nature of the state. In this section, we discuss the nonclassicality behaviour of the state by using different criteria like higher-order Mandel's $Q_M$ parameter, higher-order antibunching (HOA), higher-order sub-Poissonian photon statistics (HOSPS), higher-order squeezing (HOS) of Hong-Mandel type, Aggarwal-Tara and Klyshko's conditions.

Since most of these experimentally measurable nonclassicality witnesses can be expressed in terms of the moments of annihilation and creation operators \cite{allevi}, it is beneficial to find out an analytic expression for the most general moment $\braket{a^{\dagger m}a^n}$, $m$, $n$ being non-negative integers. For calculating $\braket{a^{\dagger m}a^n}$, we proceed as in follows:

\begin{eqnarray}
\begin{array}{rcl}
\label{eq2}
a a^{\dagger p} & = & aa^\dagger a^{\dagger{p-1}}\\
& = & a^{\dagger{p-1}}+a^\dagger a a^{\dagger{p-1}}\\
& = & a^{\dagger{p-1}}+a^\dagger aa^\dagger a^{\dagger{p-2}}\\
& = & 2a^{\dagger{p-1}}+a^{\dagger 2} a a^{\dagger{p-2}}\\
& = & \ldots\\
& = & p a^{\dagger{p-1}}+a^{\dagger p} a\,\,\,\mbox{(proceeding similarly $p$ times)}
\end{array}
\end{eqnarray}
Similarly we have,
\begin{eqnarray}
\label{eq3}
a^{p+1} a^{\dagger} = (p+1)a^p+a^\dagger a^{p+1}
\end{eqnarray}
Using (\ref{eq2}) and (\ref{eq3}), we have obtained
\begin{eqnarray}
\begin{array}{rcl}
\label{eq4}
& & a a^{\dagger p}a^p{a^\dagger}\\ & = & aa^{\dagger p}(p a^{p-1}+a^\dagger a^p)\\
& = & paa^{\dagger p}a^{p-1}+aa^{\dagger {p+1}}a^p\\
& = & p(p a^{\dagger{p-1}}+a^{\dagger p} a)a^{p-1}
+\left((p+1) a^{\dagger p}+a^{\dagger{p+1}} a\right)a^p\\
& = &  p^2 a^{\dagger {p-1}}a^{p-1} + (2p+1)a^{\dagger p}a^p + a^{\dagger {p+1}}a^{p+1}
\end{array}
\end{eqnarray}
Again using (\ref{eq2}), (\ref{eq3}) and (\ref{eq4}), $\braket{a^{\dagger m}a^n}$ can be derived as
\begin{widetext}
\begin{equation}
\begin{array}{rcl}
\label{eq5}
\braket{a^{\dagger m}a^n}
& = & \braket{\psi|a^{\dagger m}a^n|\psi}\\
& = & N^{-1}\braket{\alpha|\Big\{t^2 a^{\dagger {m+1}}a^{n+1}+r^2 a a^{\dagger {m}}a^{n} a^\dagger
+ rt\, a^{\dagger {m+1}}a^{n}a^\dagger+rt\, a a^{\dagger {m}}a^{n+1}\Big\}|\alpha}\\\nonumber
& = & N^{-1}\alpha^{* m-1}\alpha^{n-1}\Big[|\alpha|^4 + rt\Big\{(m+|\alpha|^2)\alpha^2+(n+|\alpha|^2)\alpha^{* 2}\Big\}\\
& & +r^2\Big\{mn+(m+n+1)|\alpha|^2\Big\}\Big]
\end{array}
\end{equation}
\end{widetext}
This analytic expression of $\braket{a^{\dagger m}a^n}$, $m$ and $n$ are non-negative integers, is of great help when we are calculating different moment-based witnesses of nonclassicality. Many other moments can be obtained from (\ref{eq5}) as particular cases, e.g.
\begin{enumerate}
\item If $\alpha$ is real then $\langle a^{\dagger m}a^n\rangle$ reduces to a polynomial in $\alpha$ given by $\langle a^{\dagger m}a^n\rangle = N^{-1}\Big[(2rt+1)\alpha^{m+n+2} + \Big\{r^2(m+n+1)+rt(m+n)\Big\}\alpha^{m+n} + r^2mn\,\alpha^{m+n-2}\Big]$\\
\item If $m=n=l$ (say) and $\alpha$ is complex then $\langle a^{\dagger l}a^l\rangle =N^{-1}{|\alpha|}^{2(l-1)}\Big[|\alpha|^4+r^2\Big\{l^2+(2l+1)|\alpha|^2\Big\}+rt(l+|\alpha|^2 )({\alpha}^2+{\alpha}^{* 2})\Big]$\\
\item If $m=n=1$ (say) and $\alpha$ is real then $\langle a^{\dagger m}a^n\rangle$ reduces to a polynomial given by $\langle a^{\dagger}a\rangle=N^{-1}\Big[(2rt+1)\alpha^4+(3r^2+2rt)\alpha^2+r^2\Big]$
\end{enumerate}

\subsection{Higher-order photon statistics}

The Mandel's parameter $Q_M$ \cite{mandel} illustrates the nonclassicality of a quantum state through its photon number distribution. The introductory definition of $Q_M$ can be generalized to an arbitrary order $l$ as \cite{sanjib}
\begin{eqnarray}
\label{eq6}
Q_M^{(l)} & = & \frac{\braket{(\Delta{\mathcal{N}})^l}}{\braket{a^\dagger a}}-1,
\end{eqnarray}
where $\Delta{\mathcal{N}}\,=\,a^\dagger a-\braket{a^\dagger a}$ is the dispersion in the number operator $\mathcal{N}=a^\dagger a$. Using the identity \cite{sanjib}
\begin{eqnarray*}
\braket{(\Delta{\mathcal{N}})^l} = \sum_{k=0}^l {l \choose k}(-1)^k\langle (a^\dagger a)^{l-k}\rangle{\langle a^\dagger a\rangle}^k
\end{eqnarray*}
and \cite{moya1}
\begin{equation}\nonumber
(a^\dagger a)^r = \sum_{n = 0}^r  S_r^{(n)}a^{\dagger n}a^n,
\end{equation}
where $S_r^{(n)}$ is the Stirling number of second kind \cite{stegun}
\begin{equation}\nonumber
S_r^{(n)} = \frac{1}{n!}\sum_{j=0}^n (-1)^{n-j}{n\choose j}j^r,
\end{equation}
the higher-order Mandel parameter $Q_M^{(l)}$ can be evaluated explicitly upto order $l$. The negativity of $Q_M^{(2)}$ signifies the negativity of the conventional Mandlel's $Q_M$. All expectations in (\ref{eq6}) have been calculated with help of (\ref{eq5}).

\begin{figure*}[ht]
\centering
\includegraphics[width=5.5cm]{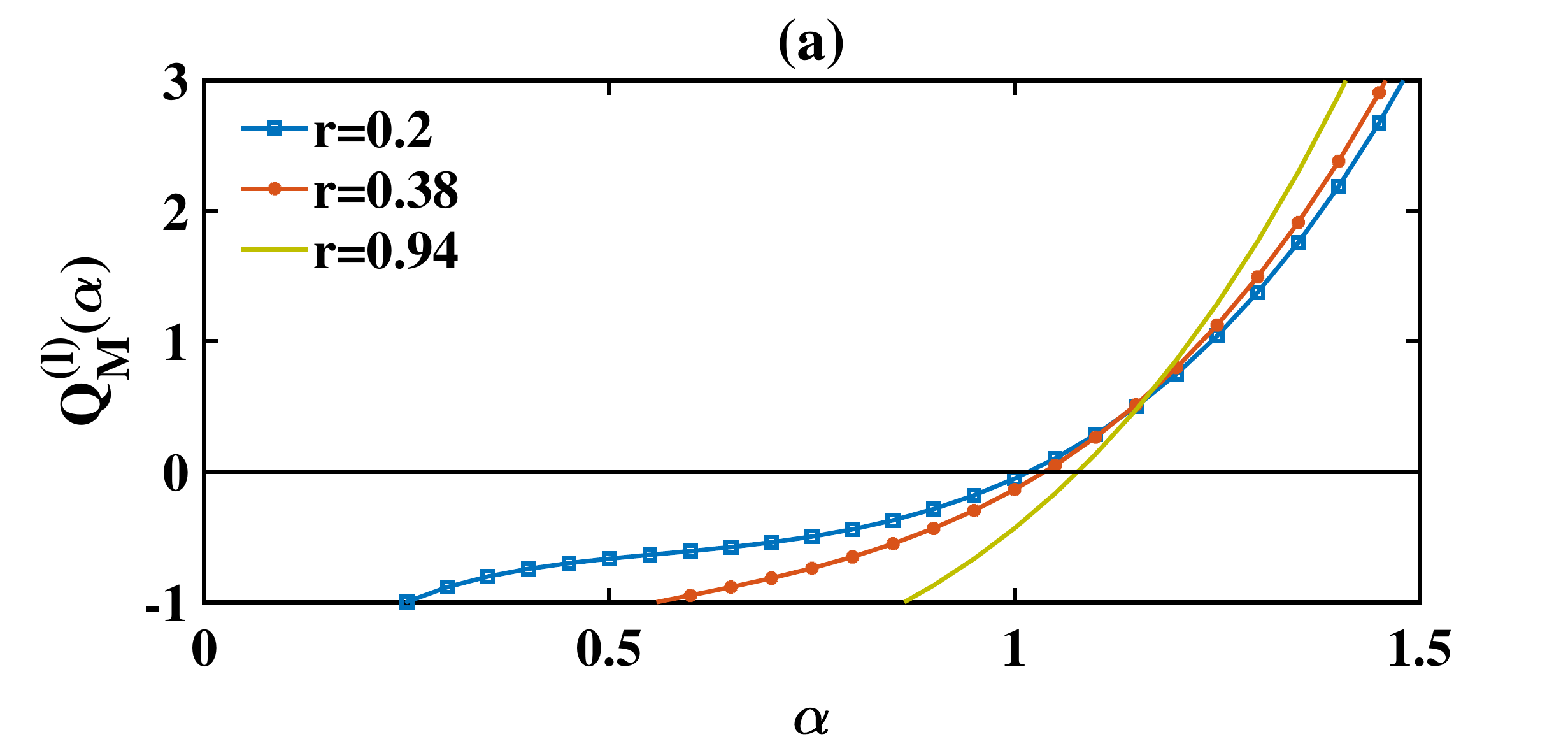}
\includegraphics[width=5.5cm]{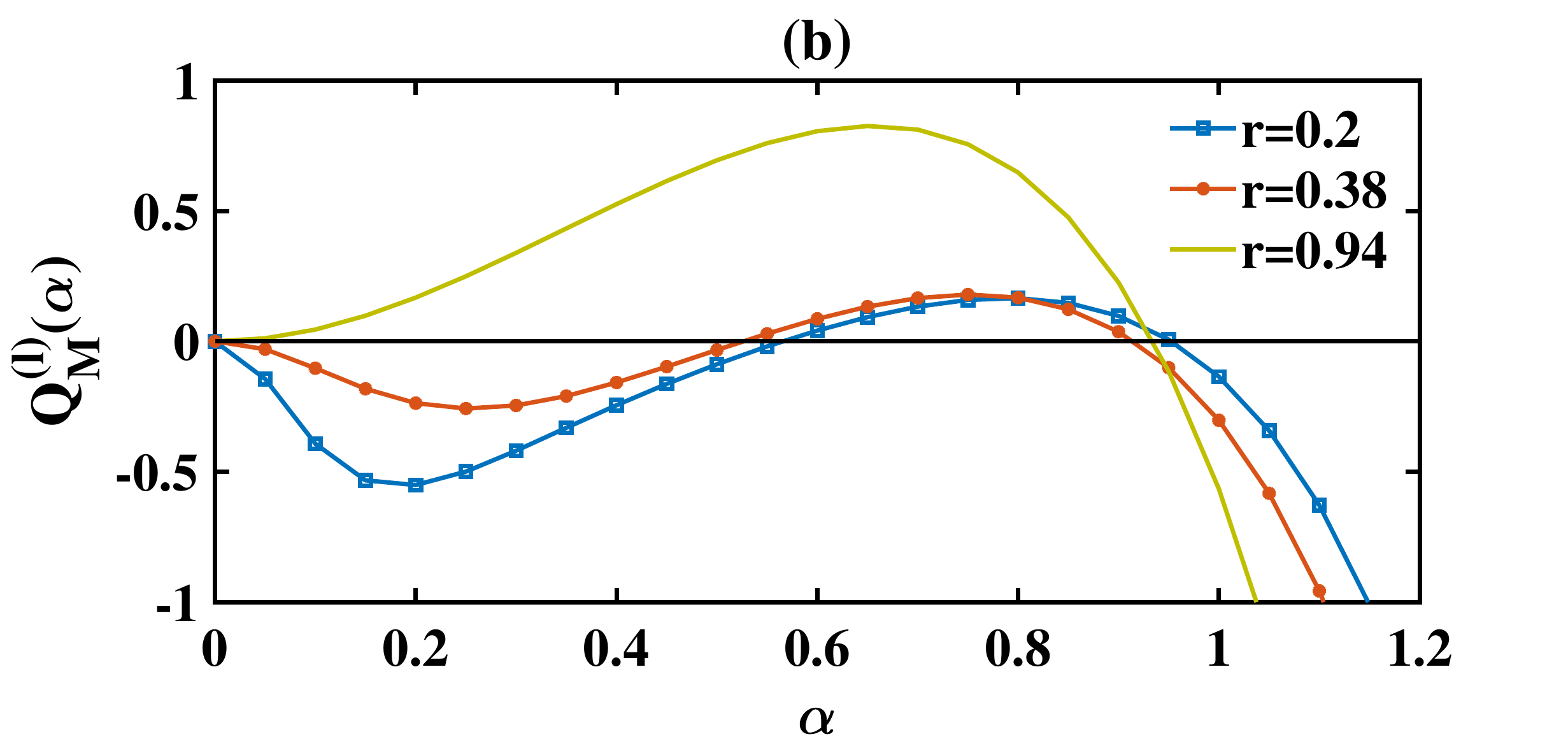}
\includegraphics[width=5.5cm]{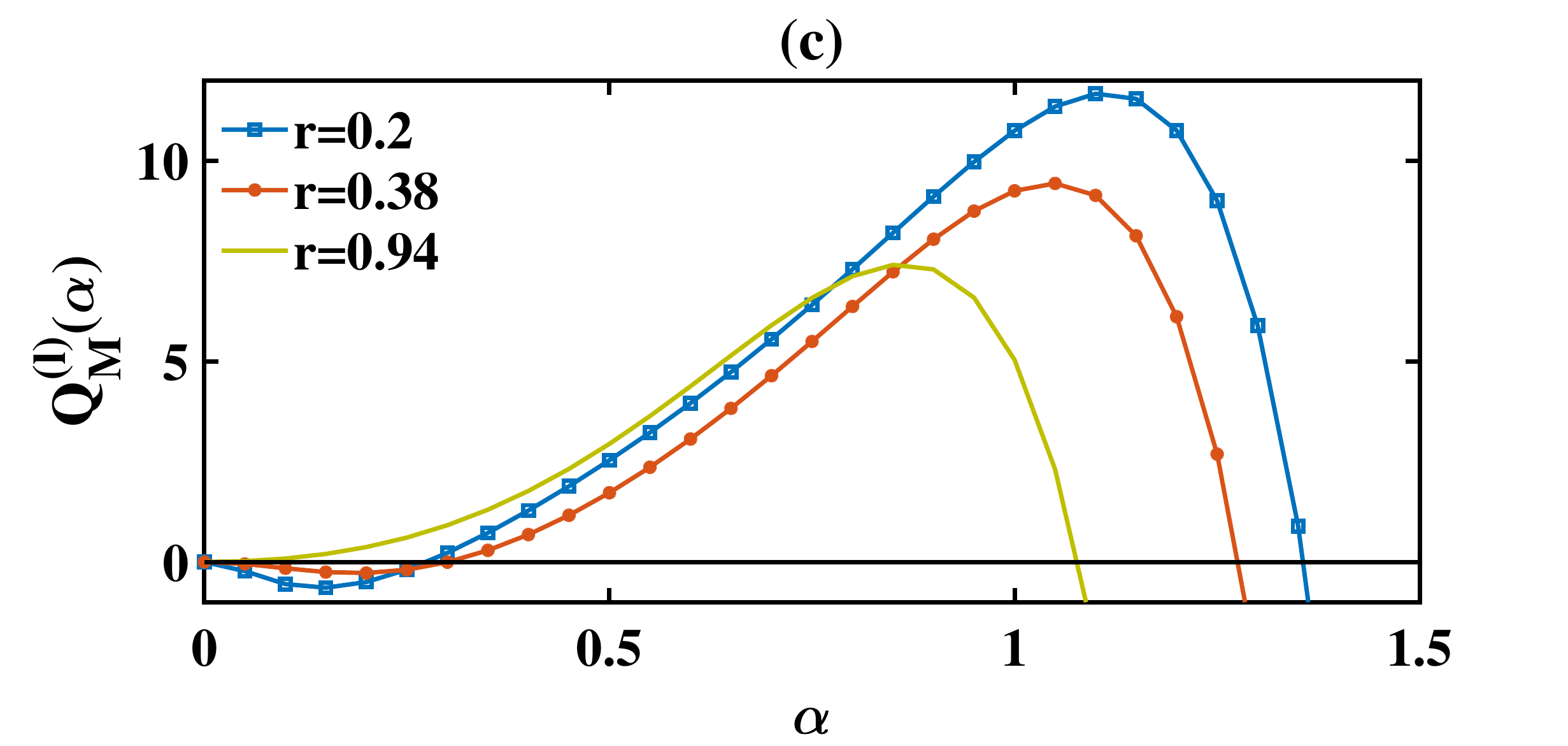}
\caption{(Color online) Comparison of $Q_M^{(l)}$ for different vales of $r$ and (a) $l=2$, (b) $l=3$ and (c) $l=5$, respectively.}
\label{fig1}
\end{figure*}

\begin{figure*}[ht]
\centering
\includegraphics[width=5.5cm]{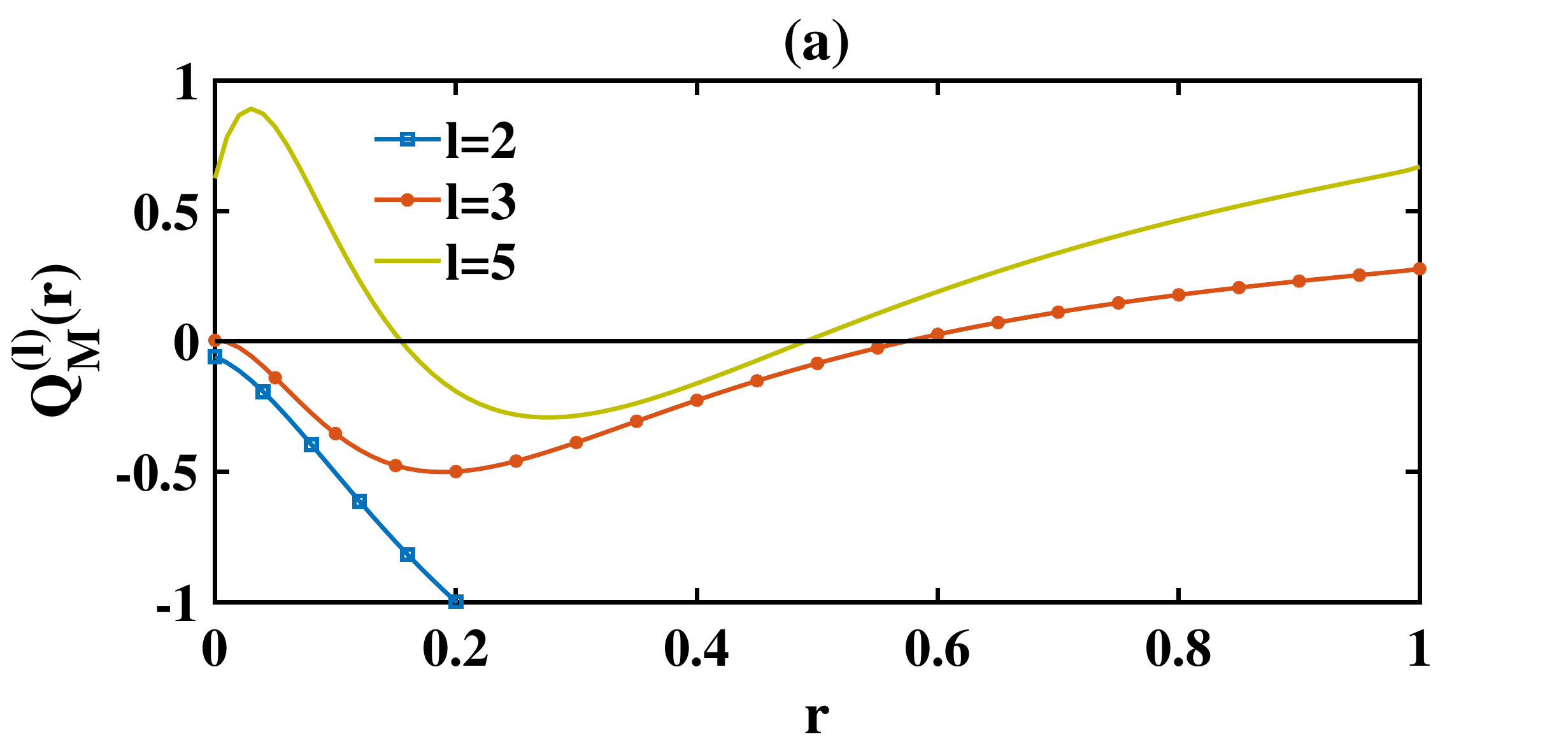}
\includegraphics[width=5.5cm]{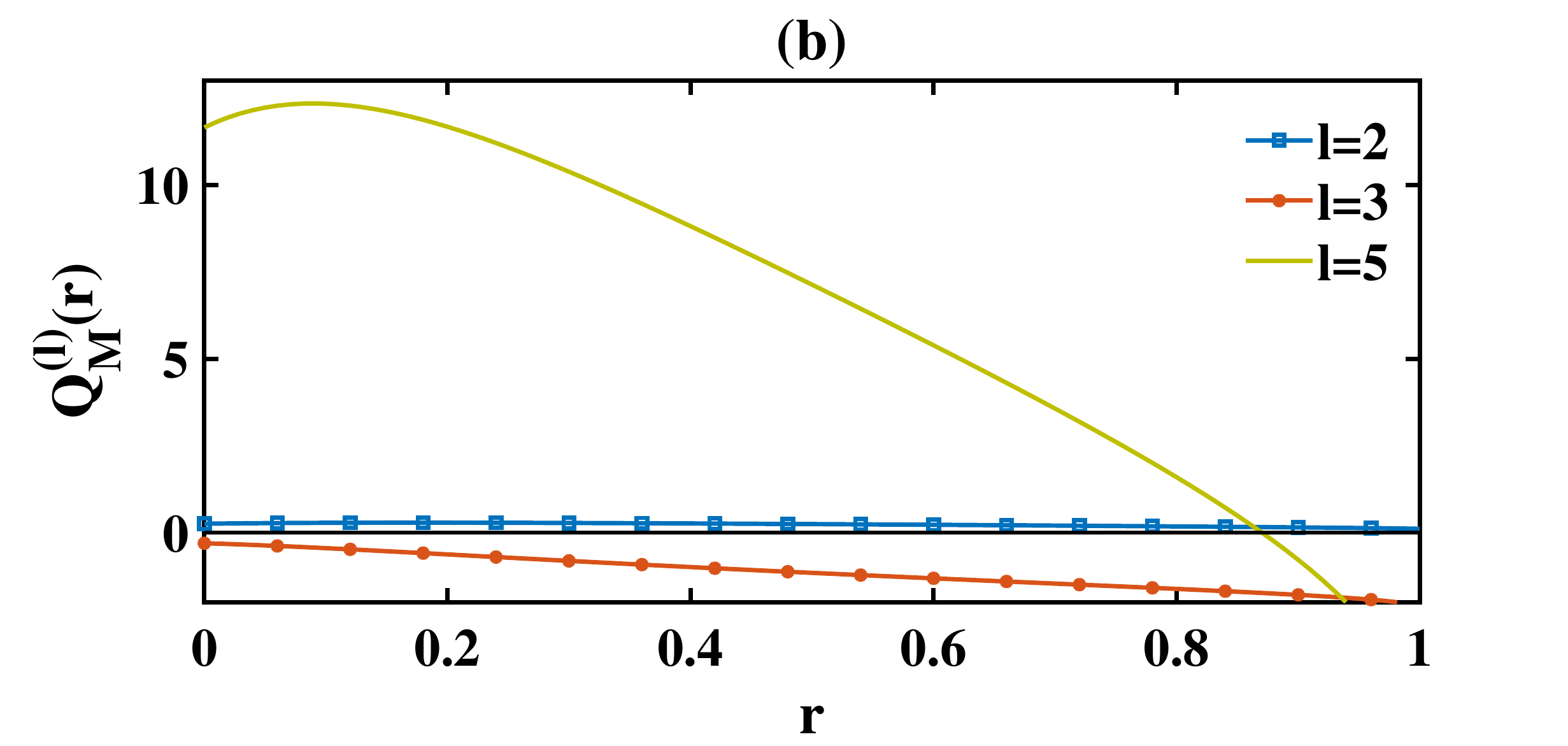}
\includegraphics[width=5.5cm]{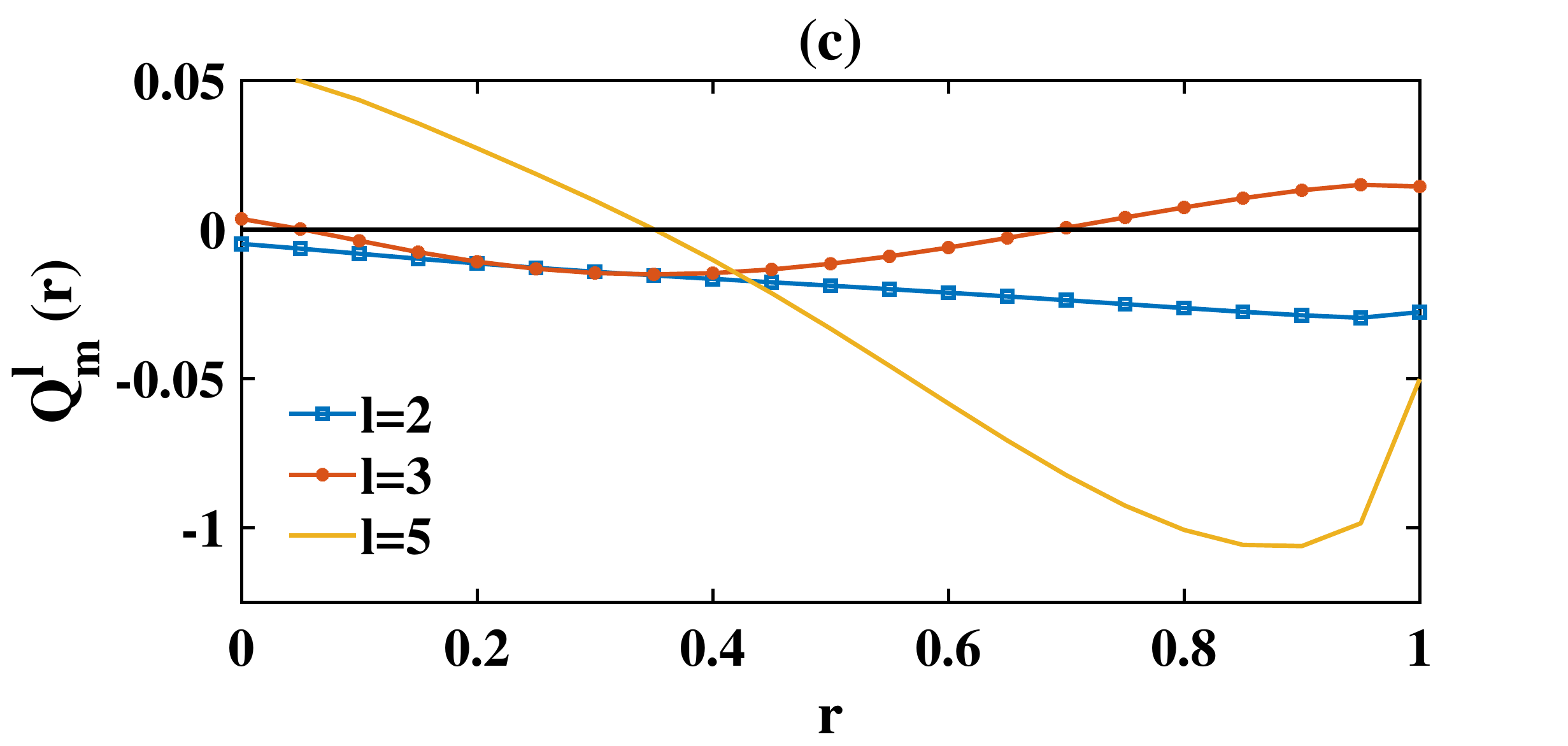}

\caption{(Color online) Plot of $Q_M^{(l)}$ as a function of $r$, for different vales of $l$ and (a) $\alpha=0.25$, (b) $\alpha=1.1$, (c) $\alpha=0.5+0.7 i$ respectively.}
\label{fig2}
\end{figure*}

The negative values of $Q_M^{(l)}$ parameter essentially indicate the negativity of the $P$ function and hence it gives a witness for nonclassicality. For all $l\geq 2$, the photon number distribution is Poissonian if $Q_M^{(l)}=0$. Whereas, $Q_M^{(l)} > 0$ and $Q_M^{(l)} < 0 $ correspond to the super-Poissonian and sub-Poissonian cases, respectively. In Fig.~\ref{fig1}, a comparison between lower- ($l=2$) and higher-order ($l=3,\,5$) Mandel's $Q_M^{(l)}$ is shown with respect to the state parameter $\alpha$ and for different values of $r$. When $l=2$ and $r=0.2,\,(0.38,\,0.94)$, the state $\ket\psi$ has $Q_M^{(l)}$ parameter value -1 corresponding to $\alpha \approx 0.25,\,(0.6,\,0.8)$, respectively, which attributes that the state becomes most nonclassical for those values. With the increase in $r$ values, the superposed state $\ket\psi$ exceeds the Poissonian limit ($Q_M^{(l)} = 0$) for larger $\alpha$. The lower-order $Q_M^{(l)}$ eventually becomes super-Poissonian if $\alpha$ increases further. While $l$ changes from 2 to 3 and then to 5 [cf. Figs.~\ref{fig1}\textcolor{blue}{(b), (c)}] and keeping $\alpha$ small ($\leq 0.4$), $\ket\psi$ initially demonstrates nonclassicality for a short range of $\alpha$. Then if $\alpha$ crosses 1, the higher-order plot has a sudden fall and $Q_M^{(l)}$ remains negative. That means the higher-order $Q_M^{(l)}$ performs better in detecting the nonclassicality and provides an enhanced sub-Poissonian characteristic for a specific choice of $\alpha$.

In Fig.~\ref{fig2}, $Q_M^{(l)}$ is plotted as a function of $r$ and for $\alpha=0.25$ and $1.1$. This figure also supports that the higher-order Mandel's $Q$ can identify the nonclassicality when $\alpha\geq 1$ but lower-order cannot. We have observed that $Q_M^{(l)}$ behaves similarly even if $\alpha$ is a complex quantity [cf. \ref{fig2}\textcolor{blue}{(c)}]. The presence of higher-order nonclassicality while its lower-order counterpart is absent approves the
relevance of the present study.

\subsection{Higher-order antibunching}

Different well-known criteria for detecting higher-order nonclassicality can be expressed in compact forms for the superposed state described in (\ref{eq1}). In this subsection, we focus on higher-order antibunching. The concept of HOA, by using the theory of majorization, was introduced by Lee \cite{ching}. Later it was modified by Pathak and Gracia \cite{gracia} to provide a clear physical meaning and a more simple expression. The $(l -1)$-th order antibunching is observed in a quantum state if it satisfies the following condition:
\begin{equation}
\label{eq7}
d(l-1)=\langle a^{\dagger l}a^l\rangle -{\langle a^\dagger a\rangle}^l\,\, <\,\,0
\end{equation}
Since the negativity of $d(l-1)$ indicates that the probability of photons coming bunched is less compared to that of coming independently, therefore the nonclassicality feature (\ref{eq7}) typifies how suitable the state $\ket\psi$ is as a single photon resource. Now
\begin{widetext}
\begin{equation}
\begin{array}{rcl}
\label{eq8}
d(l-1)& = & \braket{a^{\dagger l}a^l} -{\braket{a^\dagger a}}^l\\
& = & N^{-1}{|\alpha|}^{2(l-1)}\Big[|\alpha|^4+r^2\Big\{l^2+(2l+1)|\alpha|^2\Big\}+rt(l+|\alpha|^2 )({\alpha}^2+{\alpha}^{* 2})\Big]\\
& & -\left\{N^{-1}\Big[|\alpha|^4+r^2(1+3|\alpha|^2)+rt(l+|\alpha|^2)({\alpha}^2+{\alpha}^{* 2}\Big]\right\}^l
\end{array}
\end{equation}
\end{widetext}
The signature of lower-order antibunching can be obtained as a special case of (\ref{eq8}) for $l = 2$, and that for $l\ge 3$, the negative values of $d(l - 1)$ correspond to the higher-order antibunching of order $(l-1)$.

\begin{figure*}[ht]
\centering
\includegraphics[width=5.5cm]{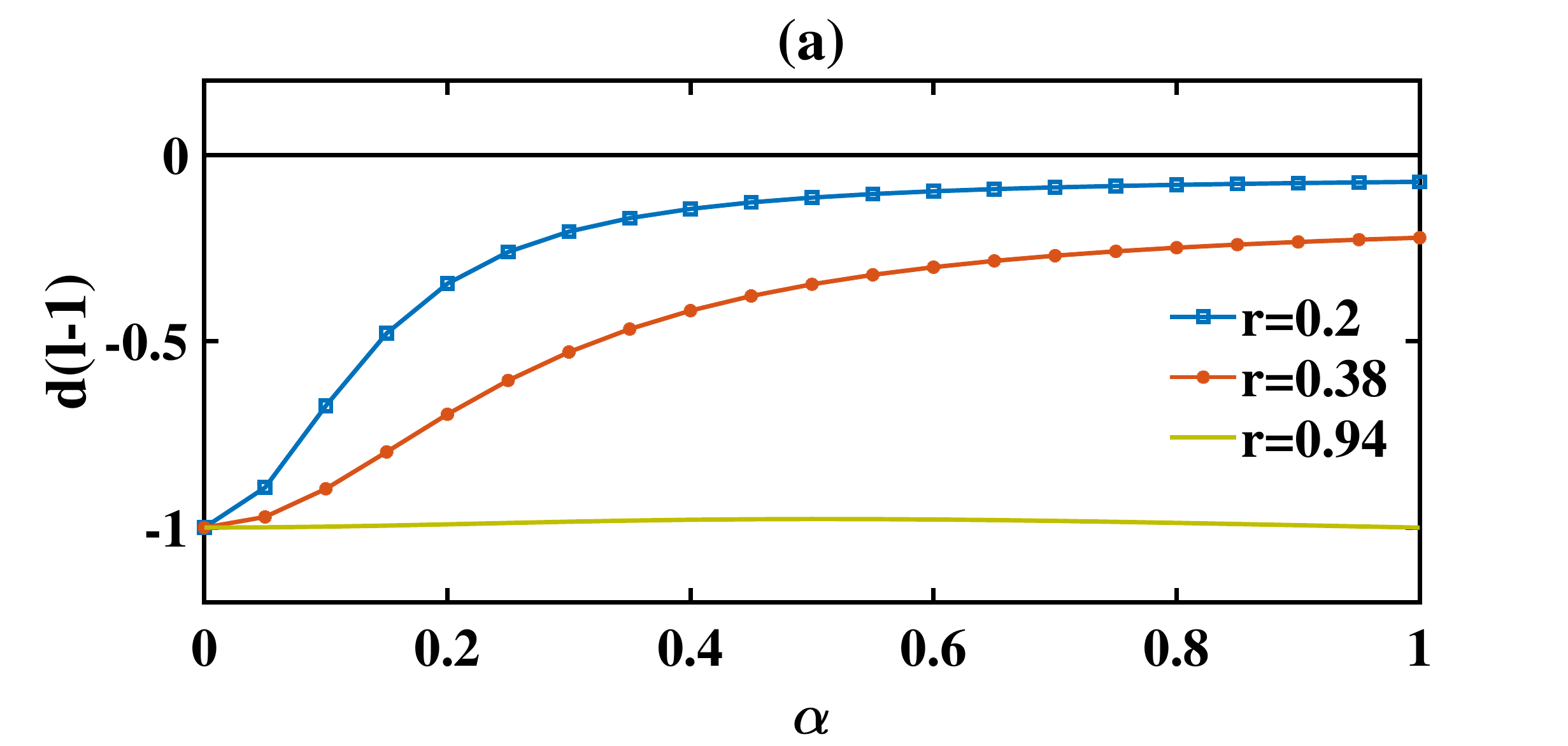}
\includegraphics[width=5.5cm]{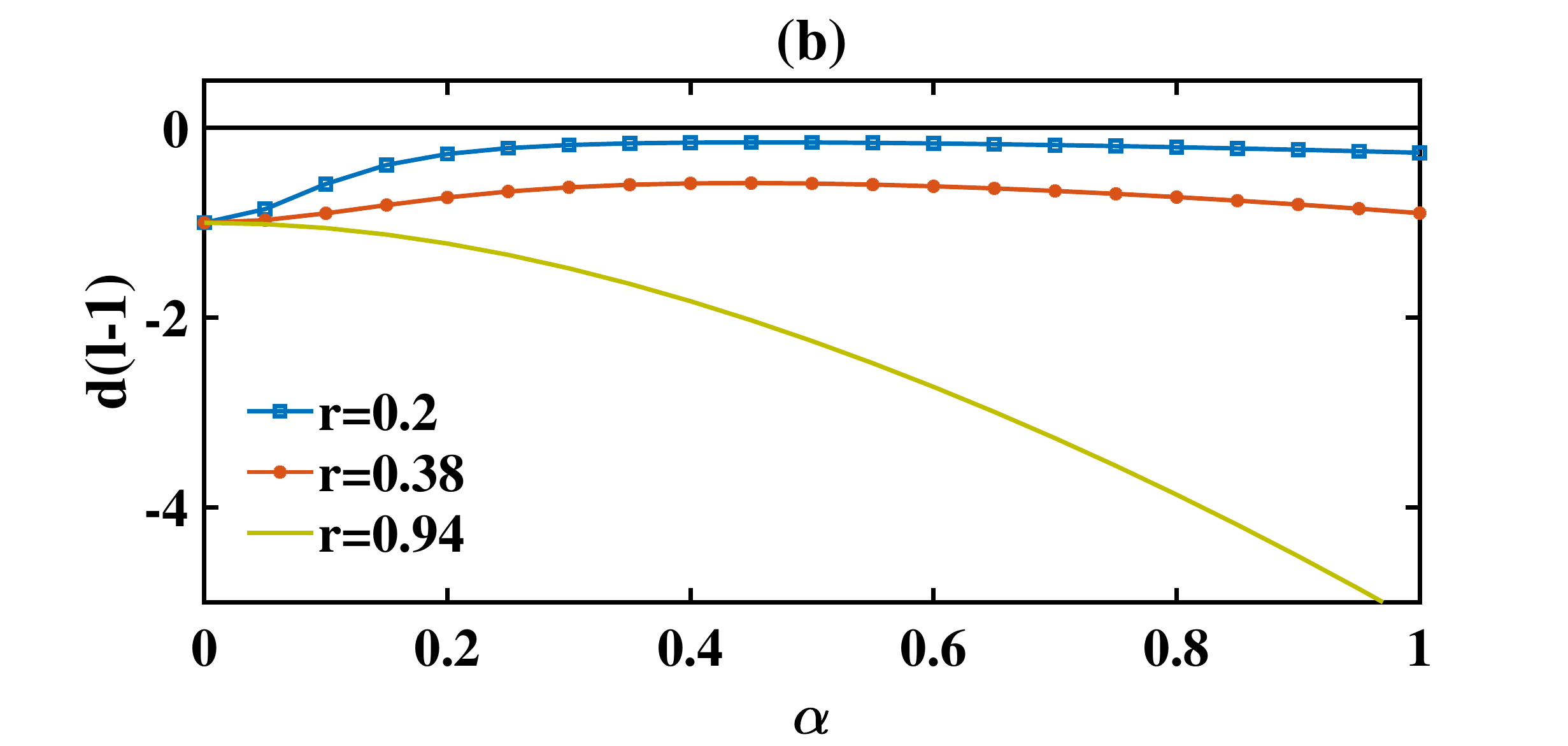}
\includegraphics[width=5.5cm]{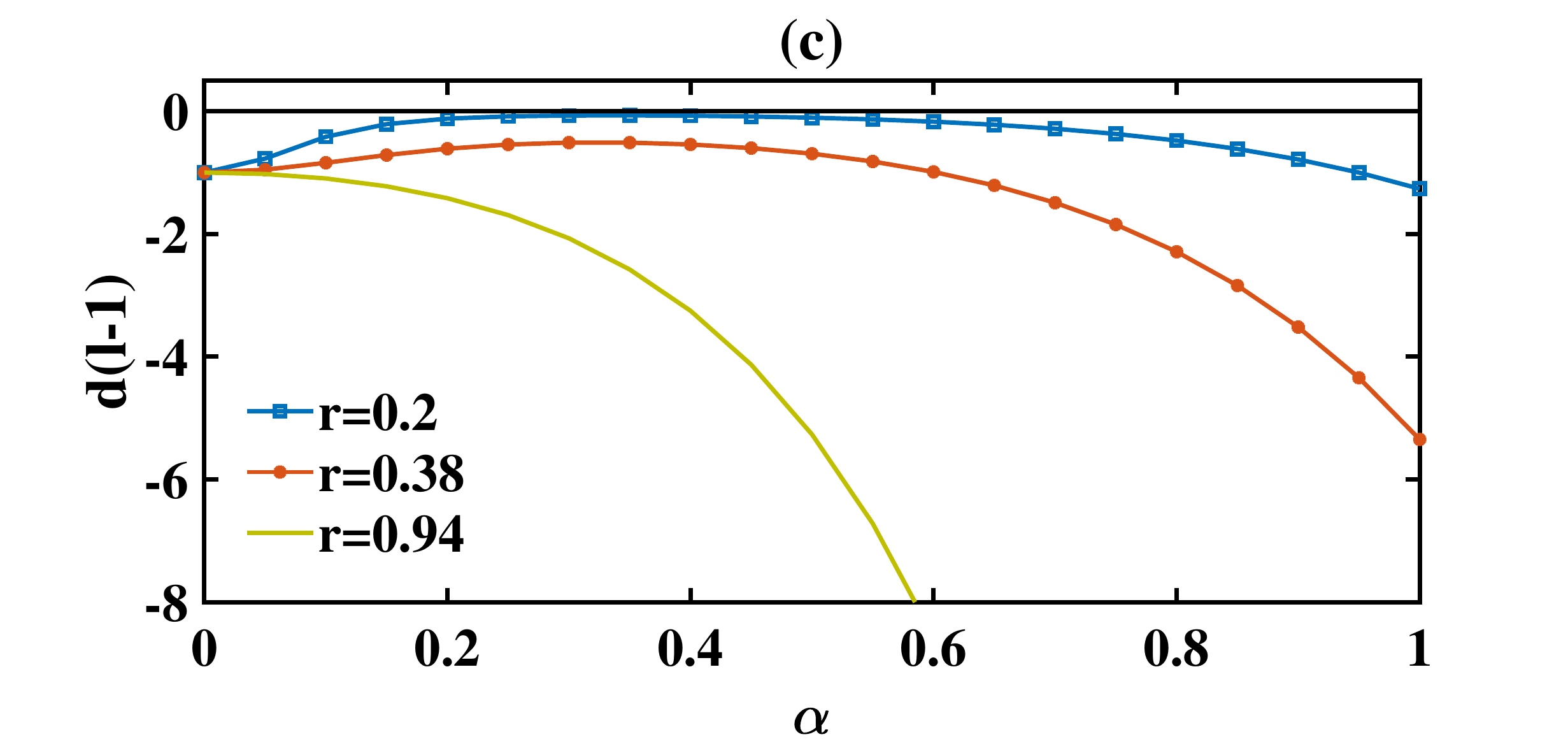}
\caption{(Color online) Comparison of $d(l-1)$ for different vales of $r$ and (a) $l=2$, (b) $l=3$ and (c) $l=5$, respectively.}
\label{fig3}
\end{figure*}

In Fig.~\ref{fig3}, the variation of lower- as well as higher-order antibuncing is shown with respect to $\alpha$. All the plots exhibit that the state is antibunched for the specific parametric values chosen here. Also Figs.~\ref{fig3}\textcolor{blue}{(b)} and \ref{fig3}\textcolor{blue}{(c)} show that the depth of nonclassicality of the superposed state $\ket\psi$ increases with the order of antibunching. This fact is consistent with the earlier observations \cite{kishore1,kishore2} that the higher-order  criteria is more effective in detecting weaker nonclassicality. It is also observed that the state is more antibunched for a relatively large value of $r$.

\subsection{Higher-order sub-Poissonian photon statistics}

Higher-order sub-Poissonian photon statistics is an important feature that affirms the existence of higher-order nonclassicality of a radiation field. The lower-order antibunching and sub-Poissonian photon statistics are closely connected as the presence of later ensures the possibility of observing the first one. But recently these two phenomena are proved to be independent of each other \cite{kishore1,kishore2}. It is also reported that the higher-order antibunching and sub-Poissonian photon statistics can exist irrespective of whether their lower-order counterparts exist or not \cite{alam1}.

The generalized criteria for observing the $(l-1)$-th order sub-Poissonian photon statistics (for which $\langle(\Delta\mathcal{N})^l\rangle < \langle(\Delta\mathcal{N})^l\rangle_{\ket{\mathrm{Poissonian}}}$ is given by \cite{amit}

\begin{equation}
\mathcal{D}_h(l-1) = \sum_{e=0}^{l}\sum_{f=1}^{e}S_2(e,f)^lC_e(-1)^ed(f-1){\langle a^\dagger a \rangle }^{l-e}\,\,<\,\,0
\label{eq9}
\end{equation}
where $S_2(e, f)=\sum_{r=0}^{f} {^fC_r}(-1)^r r^e$ is the Stirling number of second kind, $^lC_e$ is the usual binomial coefficient. The analytic expression of HOSPS for the superposed state can be obtained by substituting (\ref{eq5}) in (\ref{eq9}).
\begin{figure*}[ht]
\centering
\includegraphics[width=5.5cm]{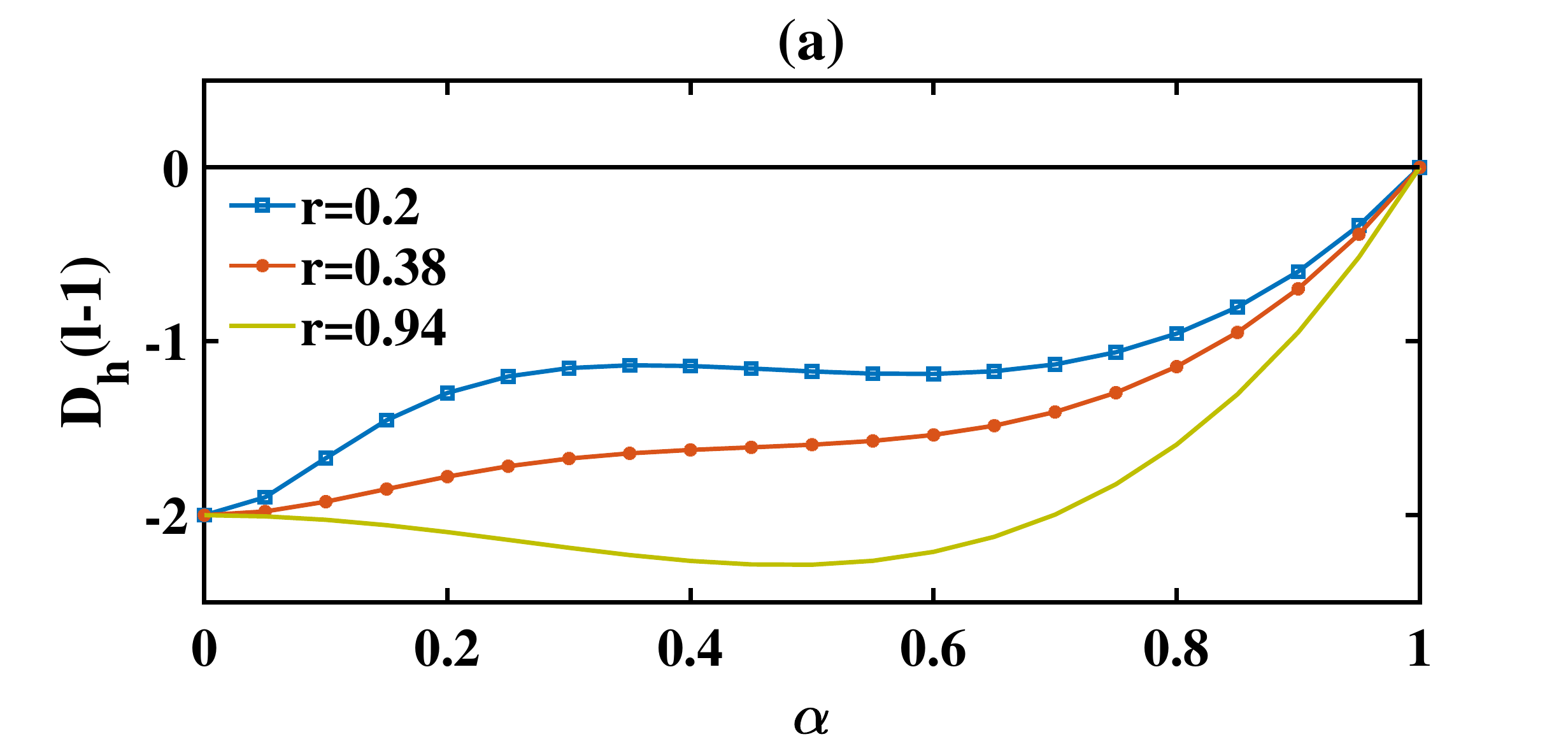}
\includegraphics[width=5.5cm]{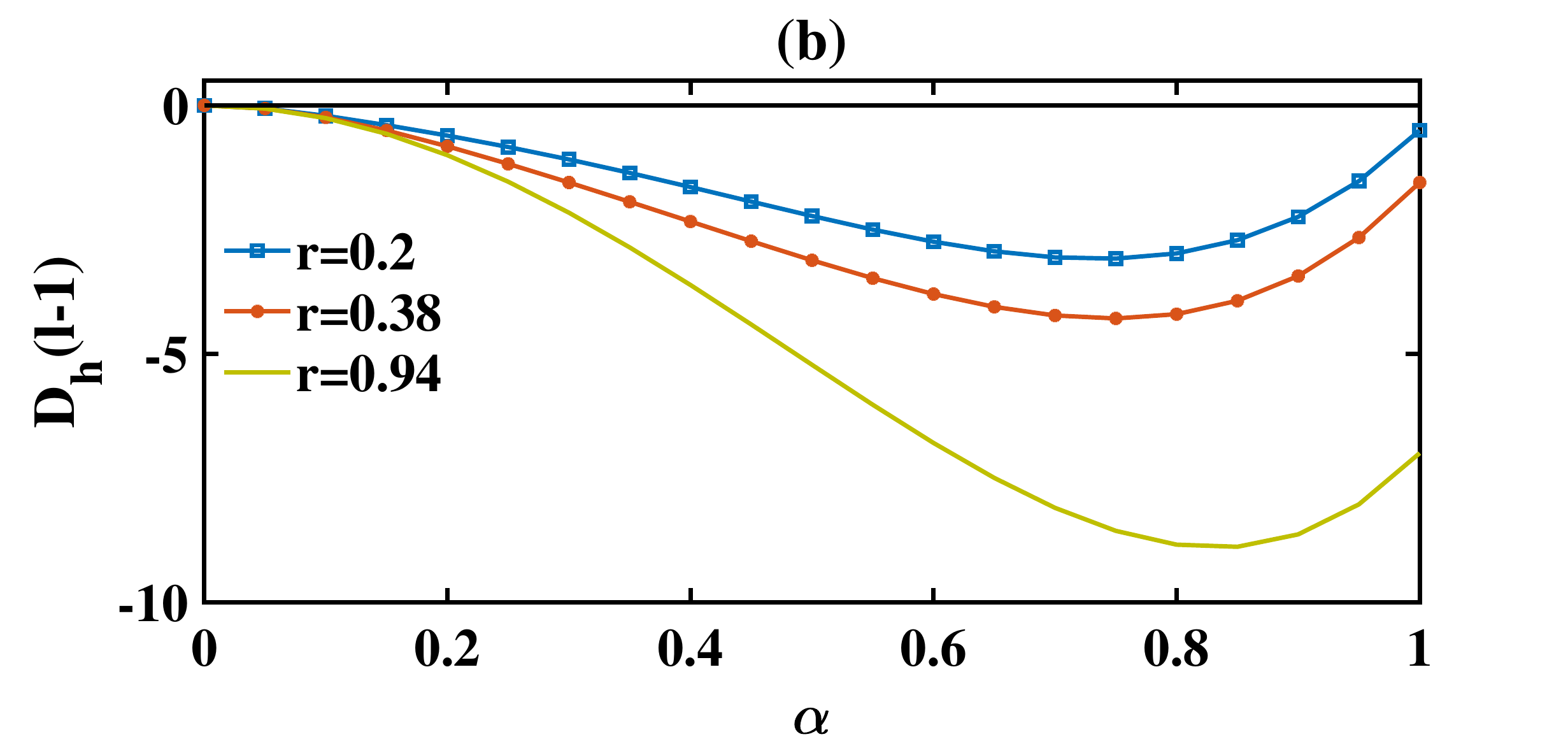}
\includegraphics[width=5.5cm]{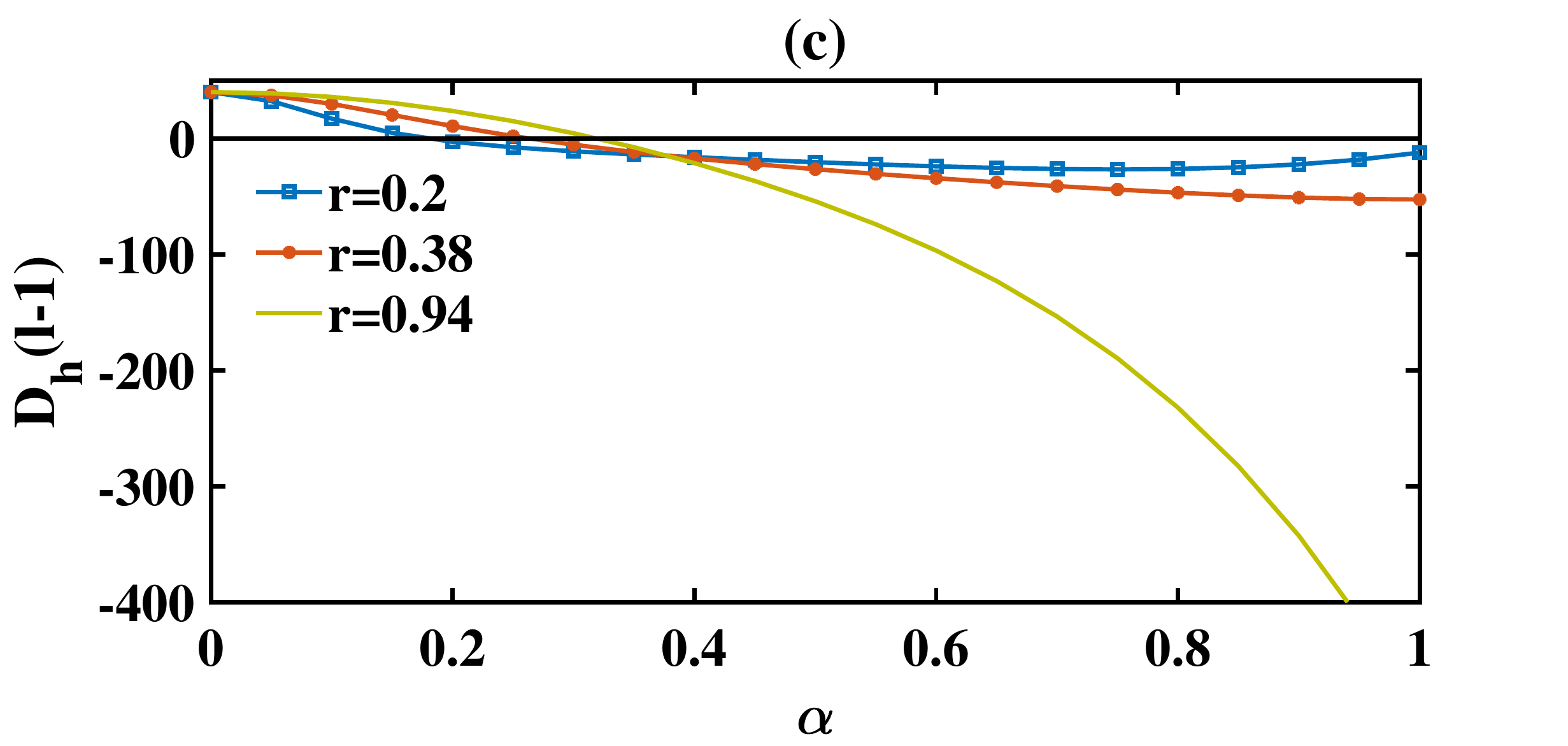}
\caption{(Color online) Comparison of $\mathcal{D}_h(l-1)$ for different vales of $r$ and (a) $l=2$, (b) $l=3$ and (c) $l=5$, respectively.}
\label{fig4}
\end{figure*}
$\mathcal{D}_h(l-1)$ is plotted in Fig.~\ref{fig4} with respect to $\alpha$ and for different values of $l$ and $r$. The figure ensures the presence of sub-Poissonian photon statistics for $l=2$ and higher-order sub-Poissonian photon statistics for $l>2$. In case of changing $l$, the behavior of HOSPS is analogous to that of HOA. That means, the depth of the nonclassicality witness increases while its order increases. Further, it can be seen that as $r$ develops from 0.2 to 0.94, lower- as well as higher-order sub-Poissonian characteristics of the superposed state $\ket\psi$ are always decreasing.

\subsection{Higher-order squeezing}

Coherent state, being the minimum uncertainty state, the product of the fluctuations in two field quadratures becomes minimum and the fluctuations in each quadrature become equal. For lower-order squeezing, the variance in one of the field quadrature (defined by a linear combination of annihilation and creation operators) reduces below the
coherent state limit at the cost of enhanced fluctuation in the other quadrature. The idea of higher-order squeezing is originated by the pioneering work of Hong and Mandel \cite{hong}. According to them, the $l$-th order higher-order squeezing ($l>2$) is obtained while the $l$-th order moment of a field quadrature operator is less than the corresponding coherent state value. Hong-Mandel's criteria for higher-order squeezing can be described by the following inequality
\begin{equation}
\label{eq10}
S(l) = \frac{\langle (\Delta X)^l \rangle -{\left(\frac{1}{2}\right)}_{\left(\frac{l}{2}\right)}}{{\left(\frac{1}{2}\right)}_{\left(\frac{l}{2}\right)}}\,\,<\,\,0,
\end{equation}
where $(x)_l$ is the conventional Pochhammer symbol and the quadrature variable is defined as $X = \frac{1}{\sqrt{2}}(a+a^\dagger)$. The inequality in (\ref{eq10}) can also be rewritten as
\begin{equation}
\label{eq11}
\langle (\Delta X)^l \rangle\,\,<\,\,{\left(\frac{1}{2}\right)}_{\left(\frac{l}{2}\right)} = \frac{1}{2^{\frac{l}{2}}}(l-1)!!,
\end{equation}
with
\begin{equation}
\begin{array}{rcl}
\label{eq12}
\langle (\Delta X)^l \rangle & = & \sum_{r=0}^{l}\sum_{i=0}^{\frac{r}{2}}\sum_{k=0}^{r-2i}(-1)^r\frac{1}{2^\frac{1}{2}}(2i-1)!^{2i}\\
& & C_k^lC_r^rC_{2i}\langle a^\dagger +a\rangle^{l-r}\langle a^{\dagger k}a^{r-2i-k}\rangle,
\end{array}
\end{equation}
where $l$ is an even number and
\begin{eqnarray*}
n!!=
\left\{
\begin{array}{lll}
& n(n-2)(n-4)\ldots 4.2 & \mbox{if $n$ is even},\\\\
& n(n-2)(n-4)\ldots3.1 & \mbox{if $n$ is odd},
\end{array}
\right.
\end{eqnarray*}
The analytic expression for the Hong-Mandel type HOS can be obtained by using (\ref{eq1}) in (\ref{eq10})-(\ref{eq12}).
\begin{figure*}[ht]
\centering
\includegraphics[width=5.5cm]{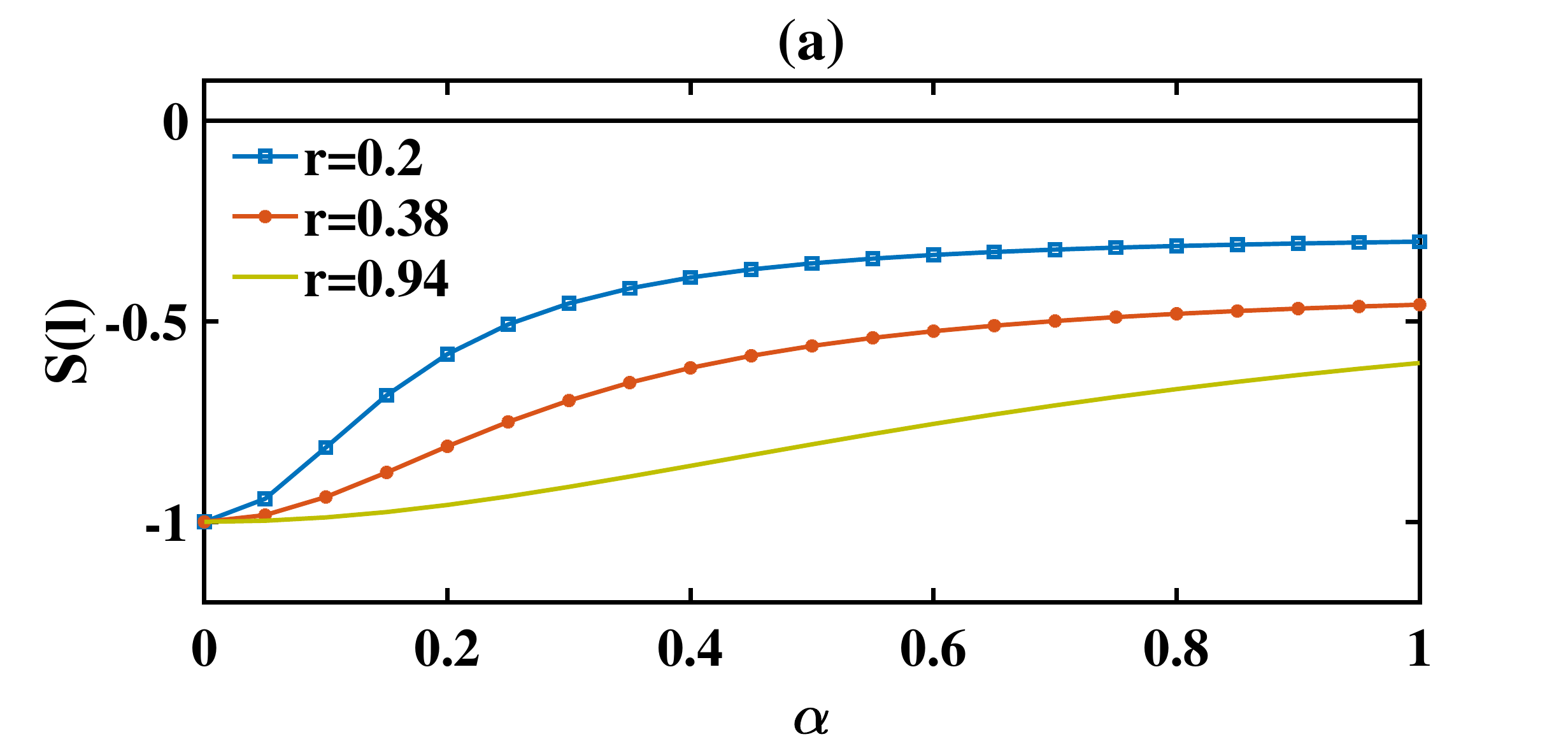}
\includegraphics[width=5.5cm]{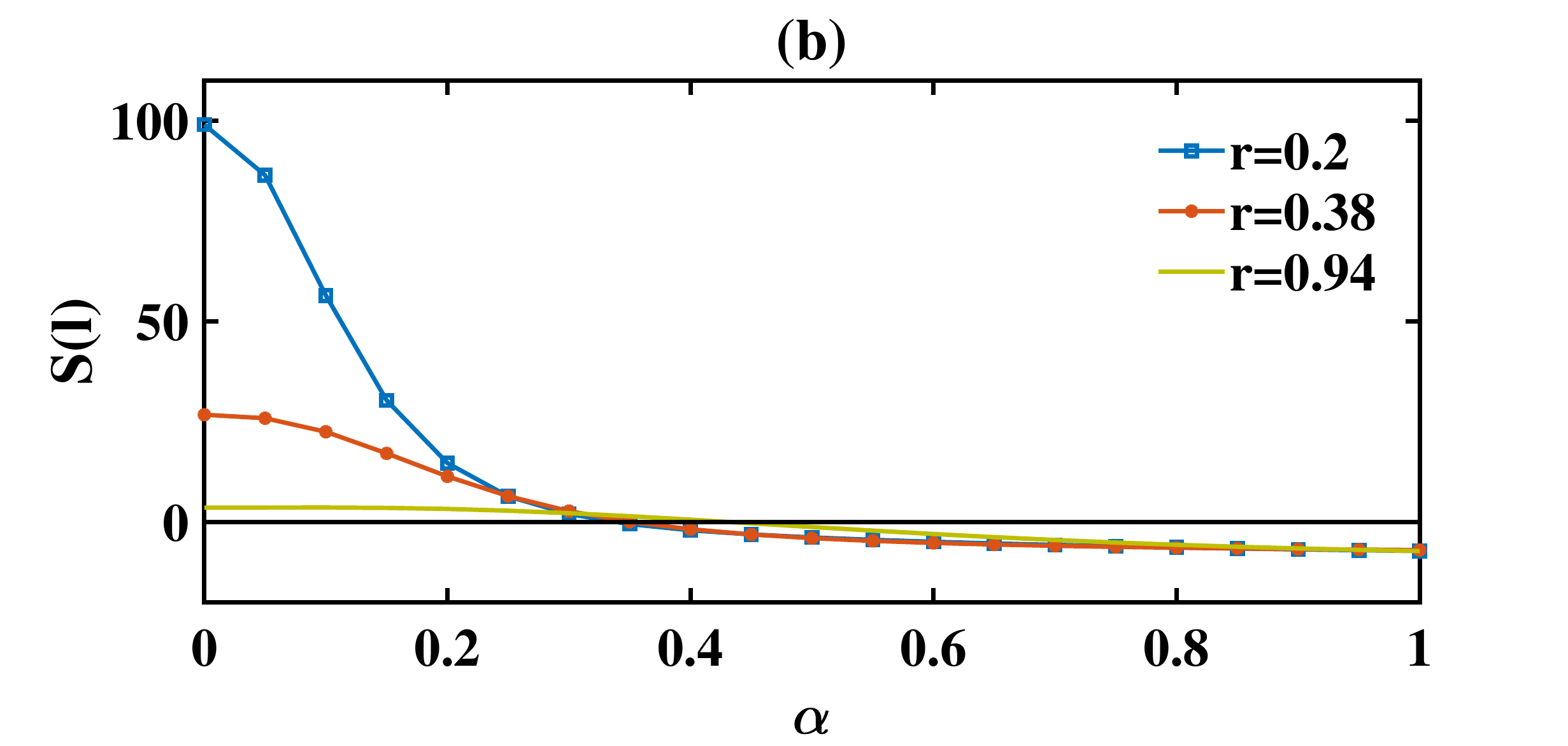}
\includegraphics[width=5.5cm]{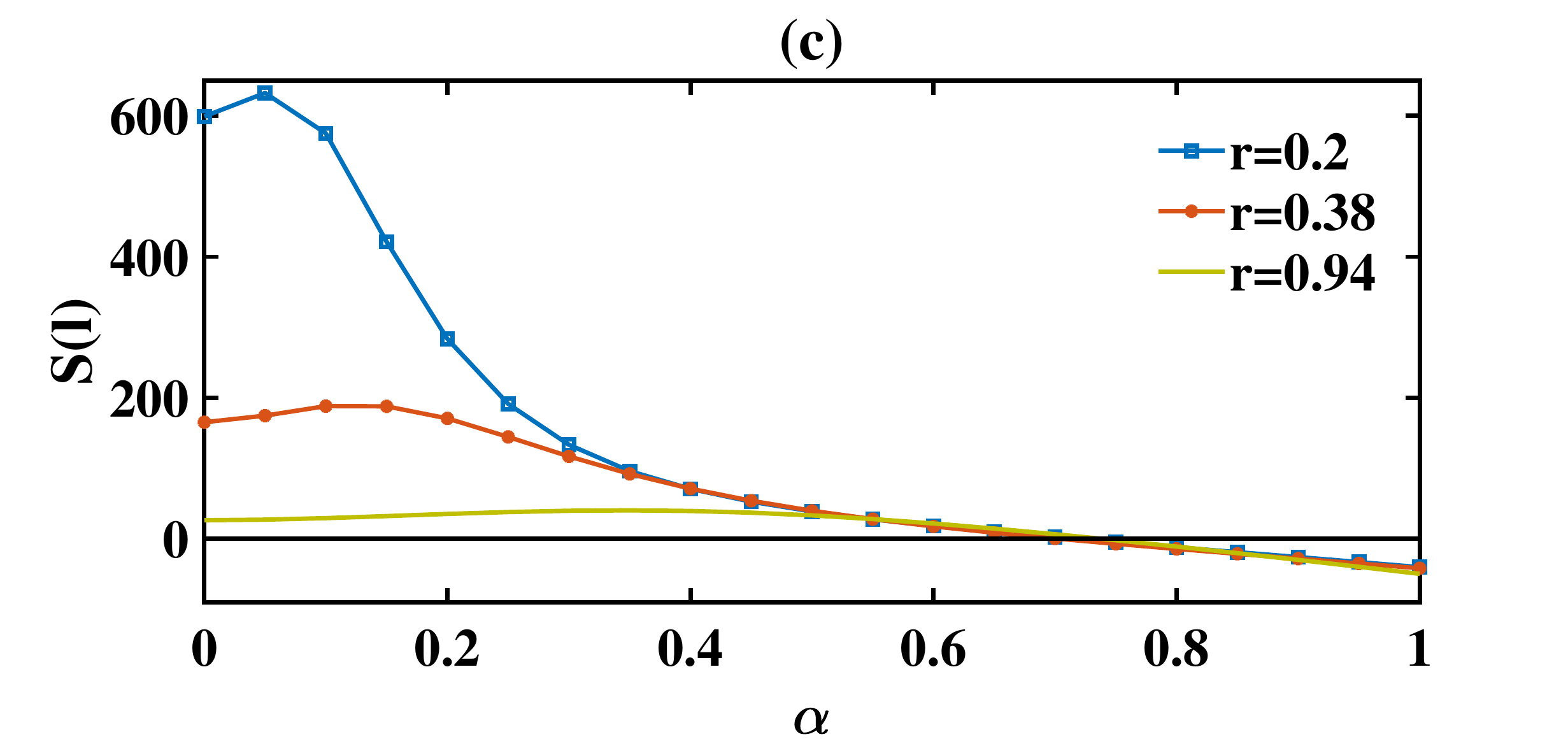}
\includegraphics[width=5.5cm]{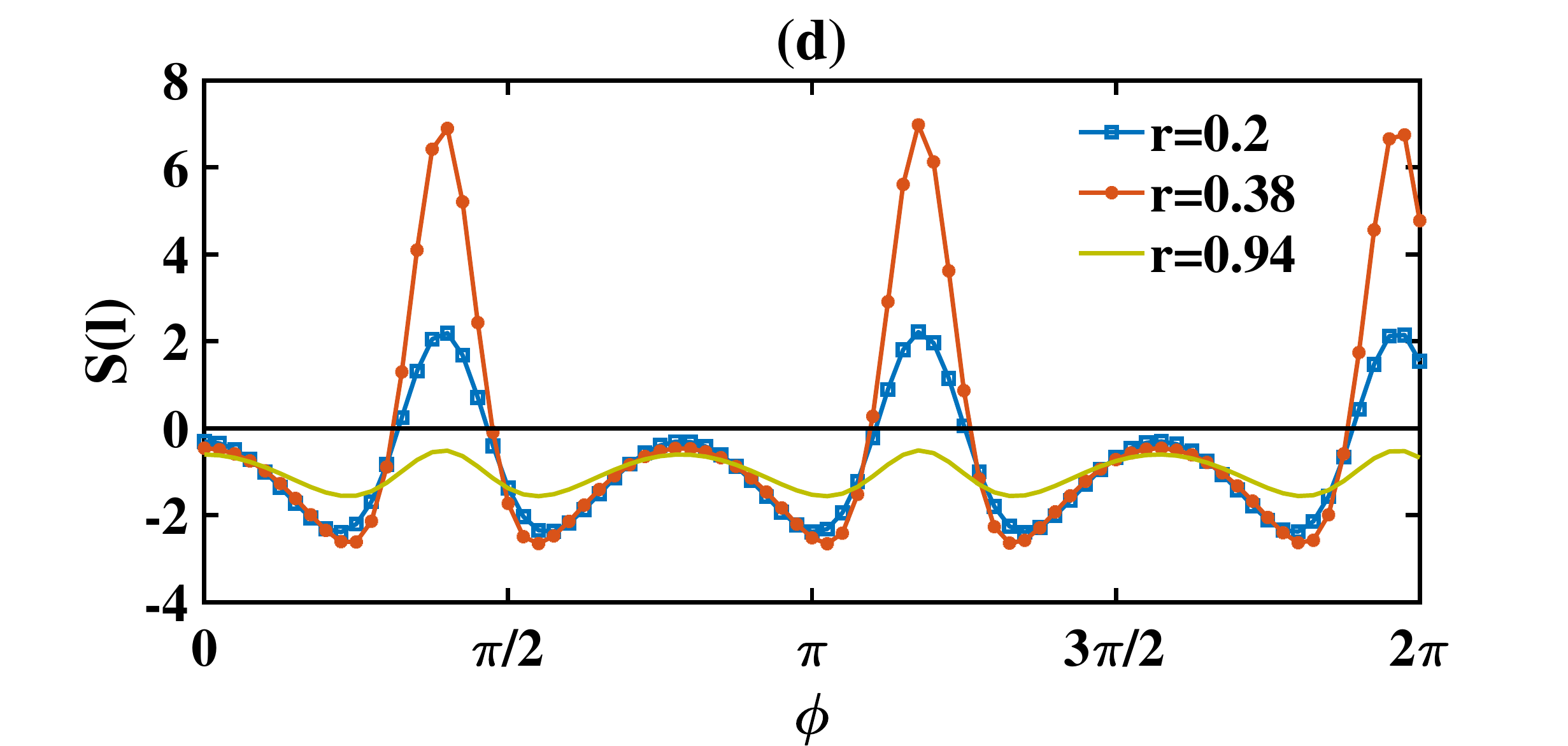}

\caption{(Color online) Hong-Mandel type higher-order squeezing $S(l)$ as a function of coherent state amplitude $\alpha$, $\alpha$ real, for different vales of $r$ and (a) $l=2$, (b) $l=4$ and (c) $l=6$, respectively, (d) lower-order squeezing as a function of phase $\phi$ of the displacement parameter $\alpha=|\alpha| e^{i\phi}$, $|\alpha|=1$.}
\label{fig5}
\end{figure*}

Fig.~\ref{fig5} illustrates the existence of Hong-Mandel type HOS in the superposed state $(t a +r a^\dagger)\ket\alpha$, assuming $\alpha$ to be real, for different orders of squeezing ($l=2,\,4,\,6$). Unlike other nonclassical features discussed so far, lower-order squeezing provides better result than the higher orders. For different values of $r$, lower-order squeezing ($S(l)$ for $l=2$) is detected throughout the range of $\alpha$. But as far as HOS is concerned, nonclassical behavior can be noticed only for higher values of $\alpha$. With increase in the order of squeezing, the state displays nonclassicality for increasing $\alpha$ further. The dependence of lower-order squeezing on the phase $\phi$ of the coherent state parameter $\alpha = |\alpha|e^{i\phi}$, taking $|\alpha|=1$, is also described here [cf. Fig.~ \ref{fig5}\textcolor{blue}{(d)}].

\subsection{$Q$ function}

A direct phase space description of a quantum mechanical system is not possible due to the uncertainty principle. This fact leads to the construction of quasiprobability distributions which are very useful in quantum mechanics as they provide a quantum classical correspondence and facilitate the calculation of quantum mechanical averages in close analogy to classical phase space averages \cite{kishore3}. One such quasiprobability distributions is $Q$ function, and zeros of this function are a signature of nonclassicality \cite{husimi}. $Q$ function is defined as
\begin{equation}\nonumber
Q = \frac{1}{\pi}\bra{\beta}\rho\ket{\beta},
\end{equation}
where $\ket{\beta}$ is the usual coherent state. This can be calculated as
\begin{eqnarray}
\begin{array}{rcl}
\label{eq13}
Q & = & \frac{1}{\pi}\braket{\beta|\rho|\beta}\\\\
& = & \frac{1}{\pi}N^{-1}|\braket{\beta|\psi}|^2 \\\\
& = & \frac{1}{\pi}N^{-1}|t\alpha+r\beta^*|^2|e^{-|\alpha|^2-|\beta|^2+\alpha\beta^*+\alpha^*\beta}
\end{array}
\end{eqnarray}
The zeros of Husimi $Q$ function in (\ref{eq13}) can be found when $t\alpha+r\beta^* = 0$, that means $r = \frac{\alpha}{\sqrt{\alpha^2+{\beta^*}^2}}$.

\begin{figure*}[ht]
\centering
\includegraphics[width=5.5cm]{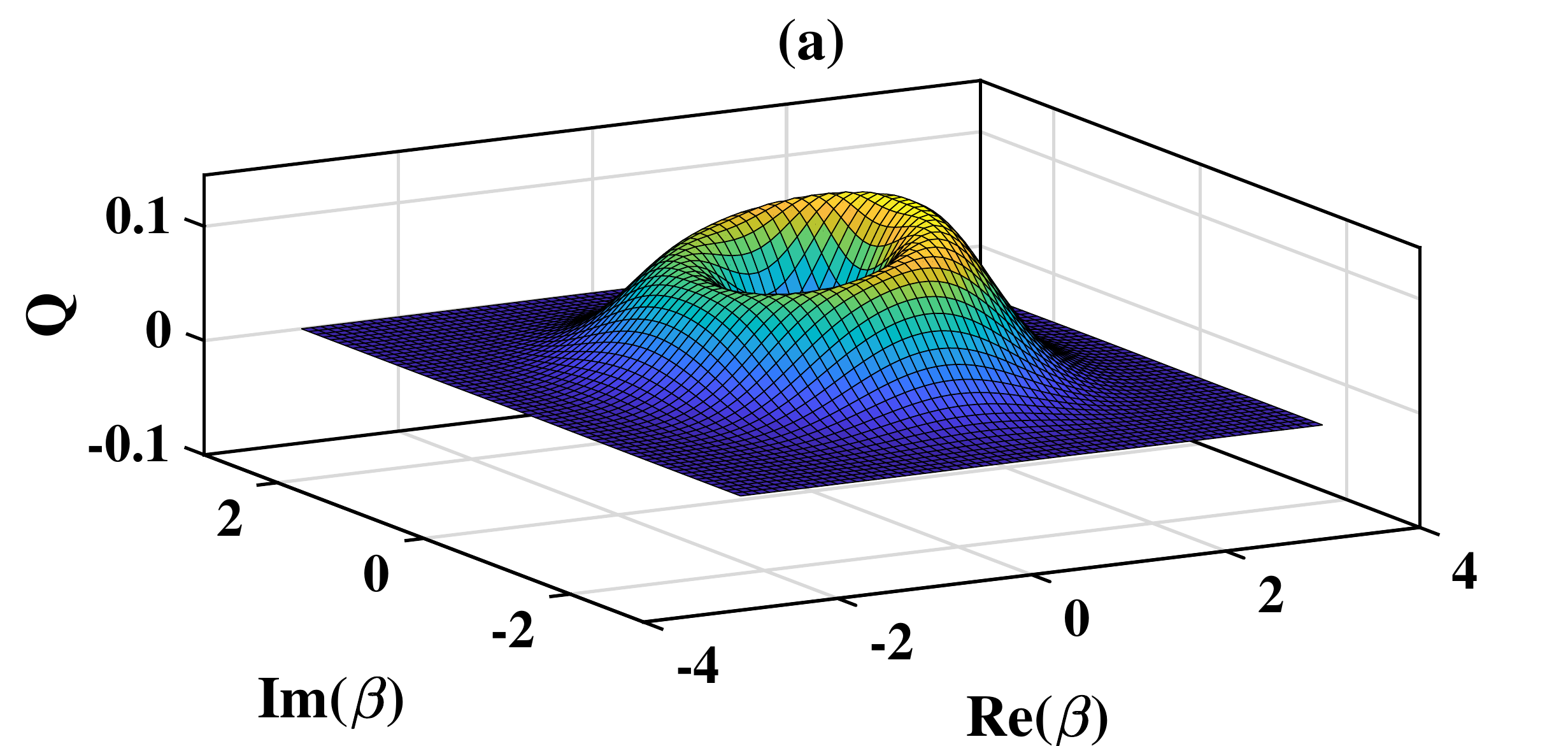}
\includegraphics[width=5.5cm]{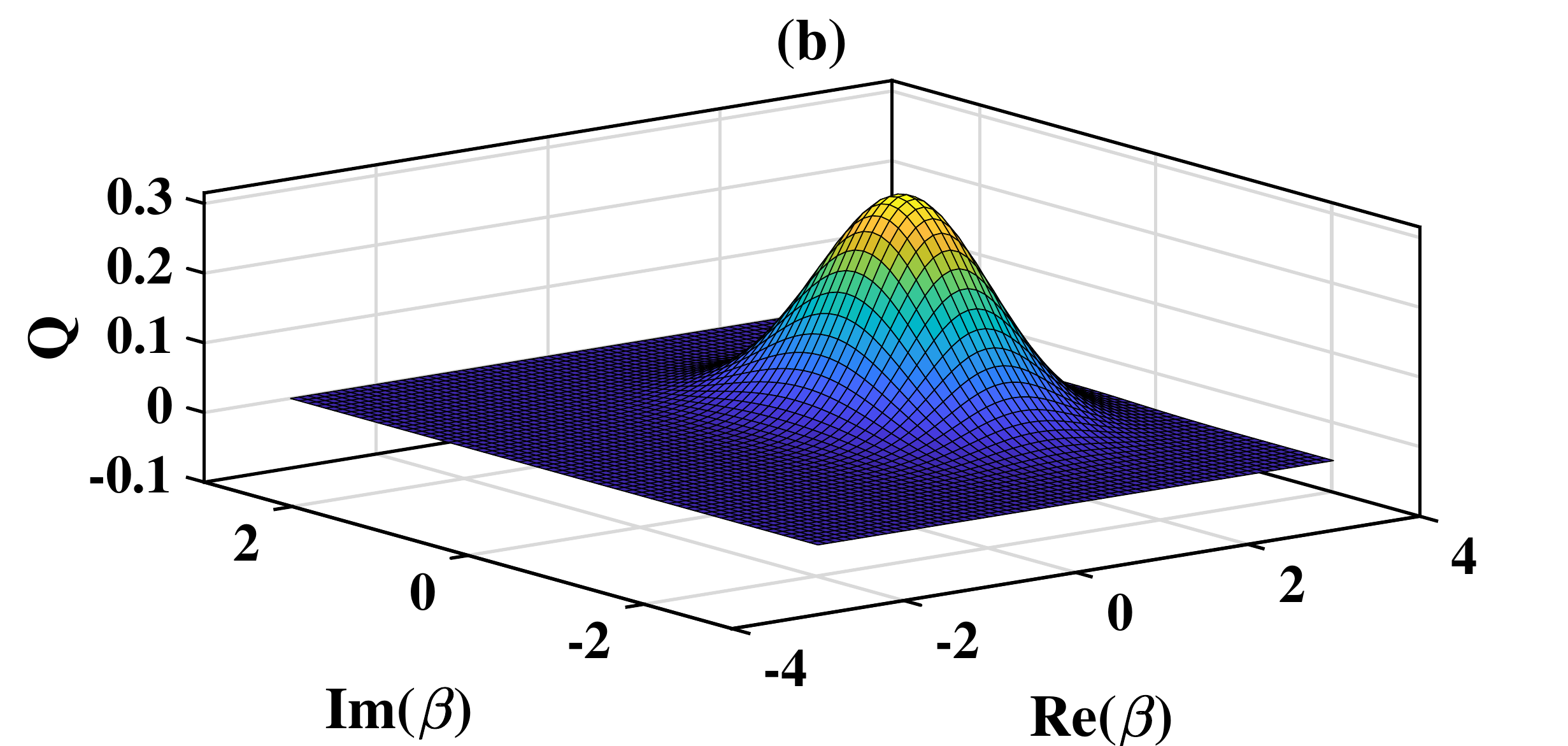}
\includegraphics[width=5.5cm]{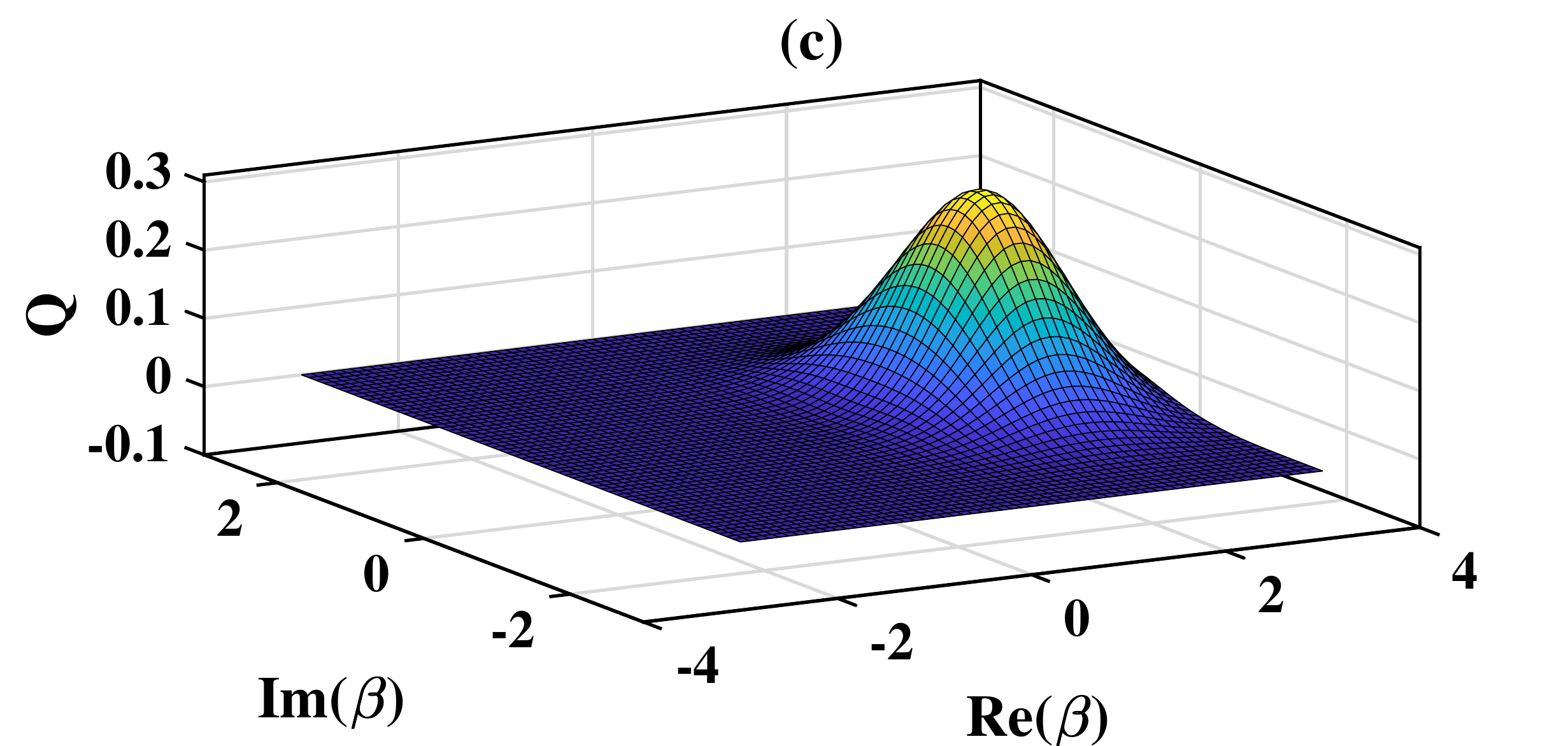}
\includegraphics[width=5.5cm]{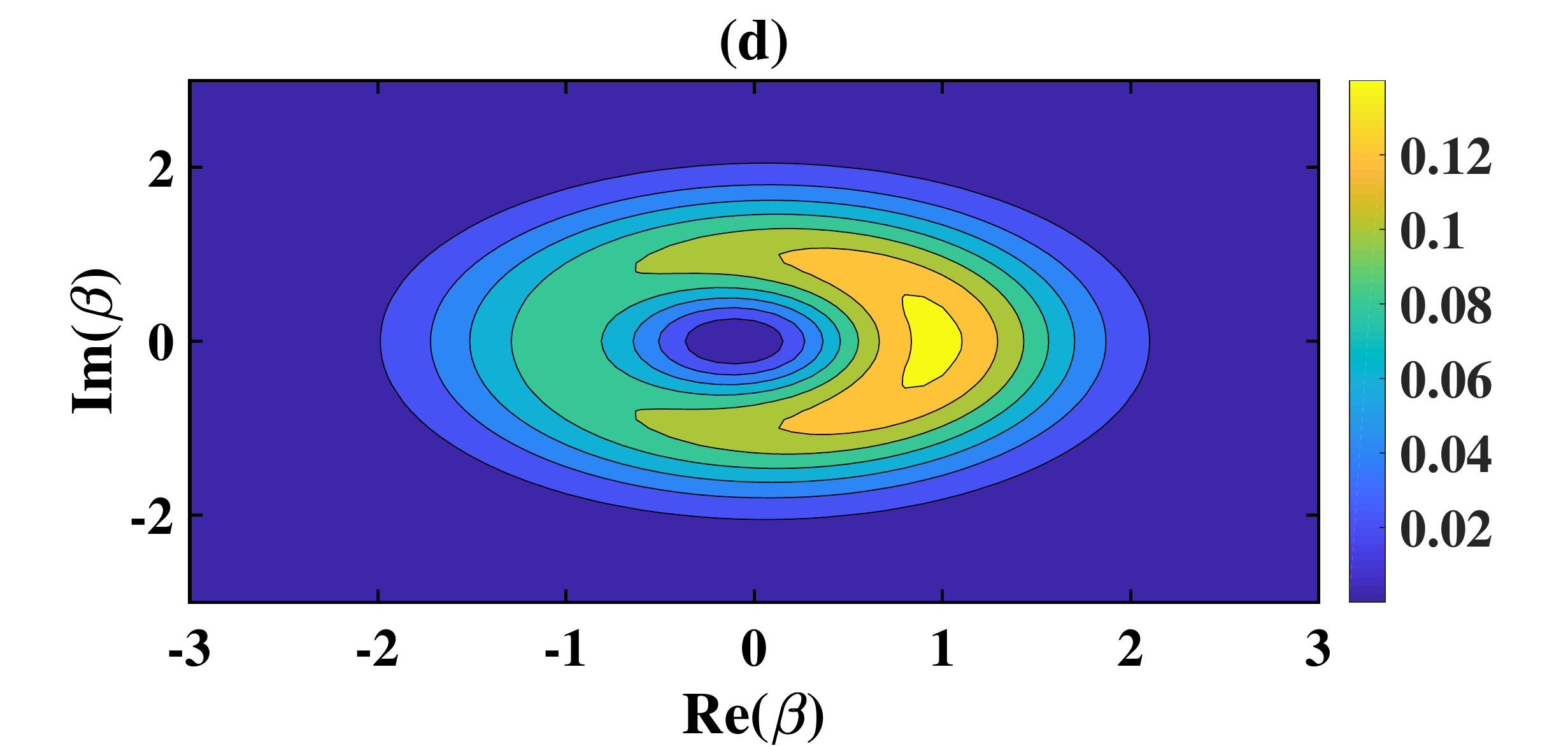}
\includegraphics[width=5.5cm]{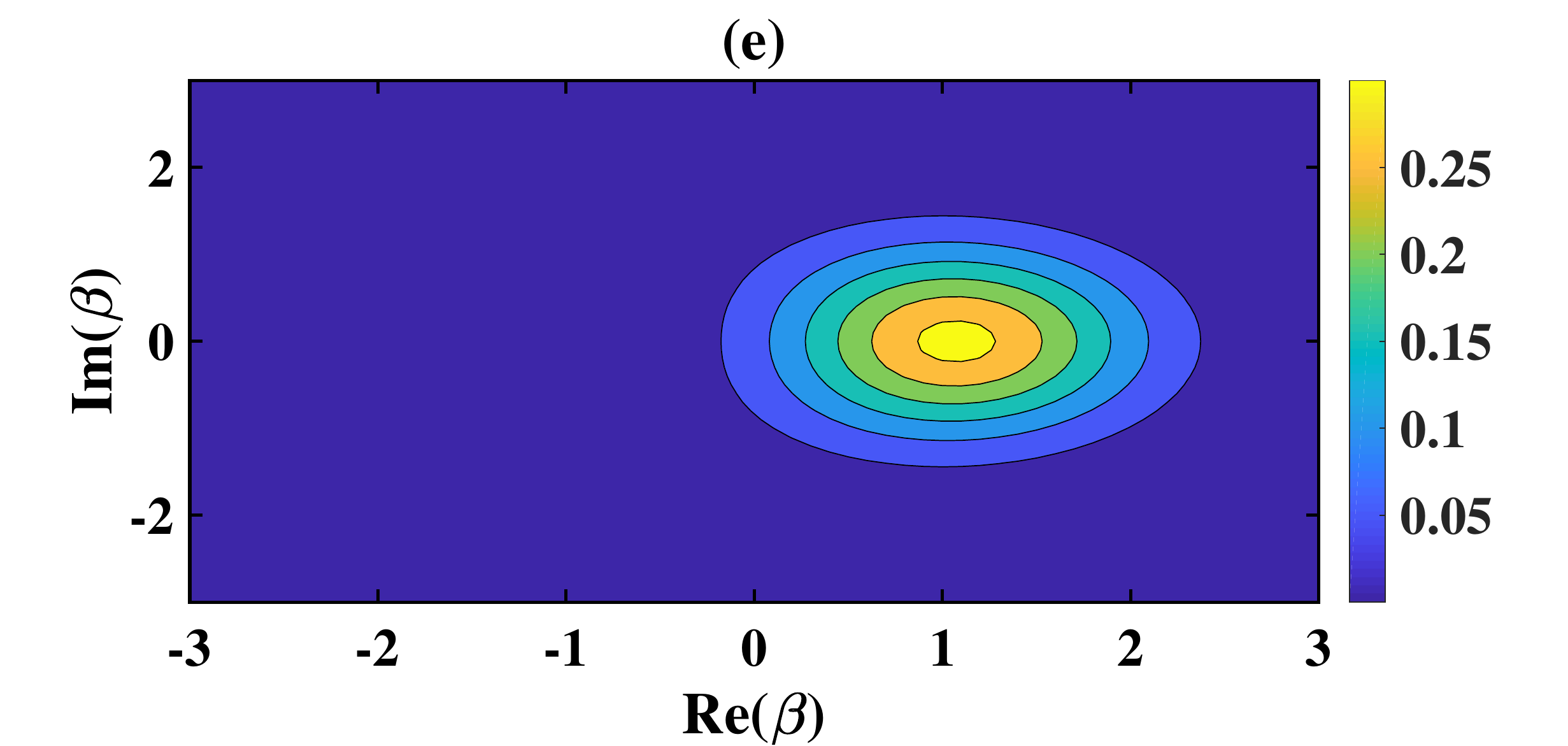}
\includegraphics[width=5.5cm]{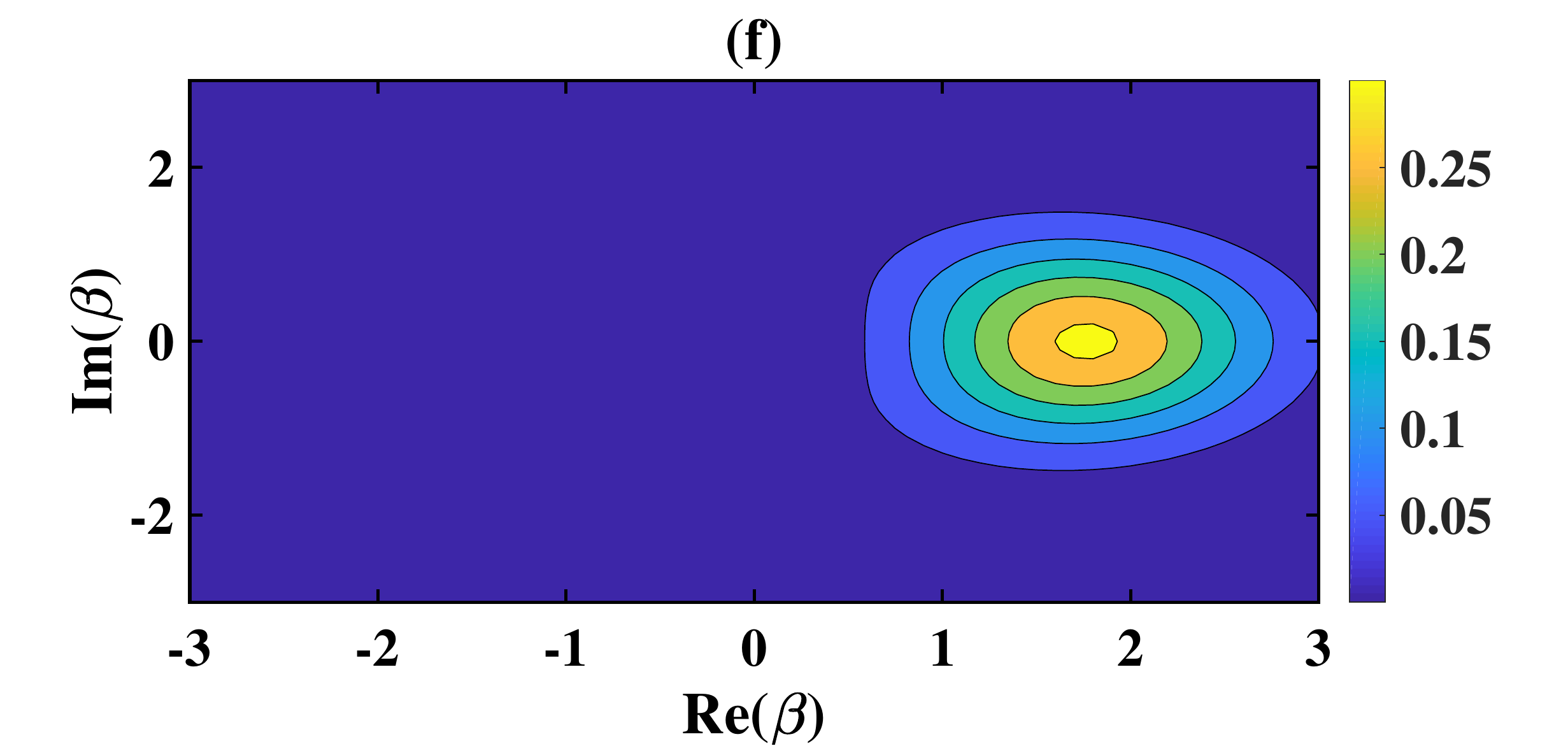}

\caption{(Color online) $Q$ as a function of $\beta$  for different vales of $\alpha$ and $r$ such as (a) $\alpha=0.02$, $r=0.2$, (b) $\alpha=0.72$, $r=0.38$ and (c) $\alpha=1.32$, $r=0.94$, respectively. Contour plots of the $Q$ function with same parametric values are given in (d), (e), (f).}
\label{fig6}
\end{figure*}

Incidentally, the quasiprobability distribution fails to grab the nonclassical features of the superposed state which are already exhibited by different moment based criteria. From Fig.~\ref{fig6}, it is understood that the values of $r$ as well as the coherent state parameter $\alpha$ have a mere effect on the Husimi's $Q$ function.

\subsection{Matrix of phase-space distributions}

Testing the nonclassical features of a physical system is a key challenge in quantum physics. Besides its fundamental importance, the notion of nonclassicality provides the basis
for many applications in photonic quantum technology and quantum information \cite{slus}. Nonclassicality is, for example, a resource in quantum networks
\cite{yadin}, quantum metrology \cite{kwon}, boson sampling \cite{shah1}, or distributed quantum computing \cite{shah2}. A very recent way of revealing nonclassical effects is by using the matrix of phase-space distributions. The condition \cite{martin}
\begin{equation}
\label{eqphase}
\mbox{det}(M) = Q(\beta_1)Q(\beta_2)-e^{-|\beta_2-\beta_1|^2/2}\,{Q\left(\frac{\beta_1+\beta_2}{2}\right)}^2\,<\,0
\end{equation}
certifies nonclassical light, when the correlations from $Q$ functions at different points in phase space fall below the classical limit zero. For the superposed state $\ket\psi$, $Q(\beta) $ is zero when $\beta^* = -\alpha \frac{t}{r}$. Assuming $\beta_1=-\alpha^* \frac{t}{r}$, the inequality (\ref{eqphase}) yields
\begin{eqnarray}
\begin{array}{rcl}
\mbox{det}(M) & = & -e^{-\frac{|\beta_2+\alpha^* \frac{t}{r}|^2}{2}}Q\left(\frac{\beta_2-\alpha^* \frac{t}{r}}{2}\right)^2\\
& = & -\frac{1}{\pi} e^{-\frac{|\beta_2+\alpha^* \frac{t}{r}|^2}{2}}|t\alpha+r\beta_2^*|^4\,e^{-2{\left|\frac{2\alpha-\beta_2+\alpha^*\frac{t}{r}}{2}\right|}^2}
\end{array}
\end{eqnarray}
Thus $\mbox{det}(M)$ is always less than zero and equals to zero iff $\frac{\beta_2-\alpha^* \frac{t}{r}}{2} = -\alpha^* \frac{t}{r}$ which gives $ \beta_2 = -\alpha^* \frac{t}{r}=\beta_1$. Thus the special case of phase-space matrix approach confirms the nonclassicality of $\ket\psi$.

\subsection{Agarwal-Tara criterion}

Agarwal and Tara introduced a moment based criterion to investigate the witness of the nonclassical characteristics of a given quantum state \cite{girish}. They defined $A_3$ which consists of the moments of the number distribution $\mu_j$ and the normal ordered moments $m_j$. The analytic expression of $A_3$ in terms of these higher ordered moments is \cite{priya}
\begin{equation}
A_3 = \frac{{\mbox{det}\,\,m}^{(3)}}{{\mbox{det}\,\,\mu}^{(3)} - {\mbox{det}\,\,m}^{(3)}}\,\, <\,\, 0,
\end{equation}
where
\begin{eqnarray*}
m^{(3)} =
\left(
\begin{array}{ccc}
1 & m_1 & m_2\\
m_1 & m_2 & m_3\\m_2 & m_3 & m_4
\end{array},
\right)
\end{eqnarray*}
and
\begin{eqnarray*}
\mu^{(3)} =
\left(
\begin{array}{ccc} 1 & \mu_1 & \mu_2\\
\mu_1 & \mu_2 & \mu_3\\\mu_2 & \mu_3 & \mu_4
\end{array}
\right)
\end{eqnarray*}
The matrix elements are defined by $m_j = \braket{a^{\dagger j}a^j}$ and  $\mu_j = \braket{(a^\dagger a)^j}$.

\begin{figure*}[ht]
\centering
\includegraphics[width=5.5cm]{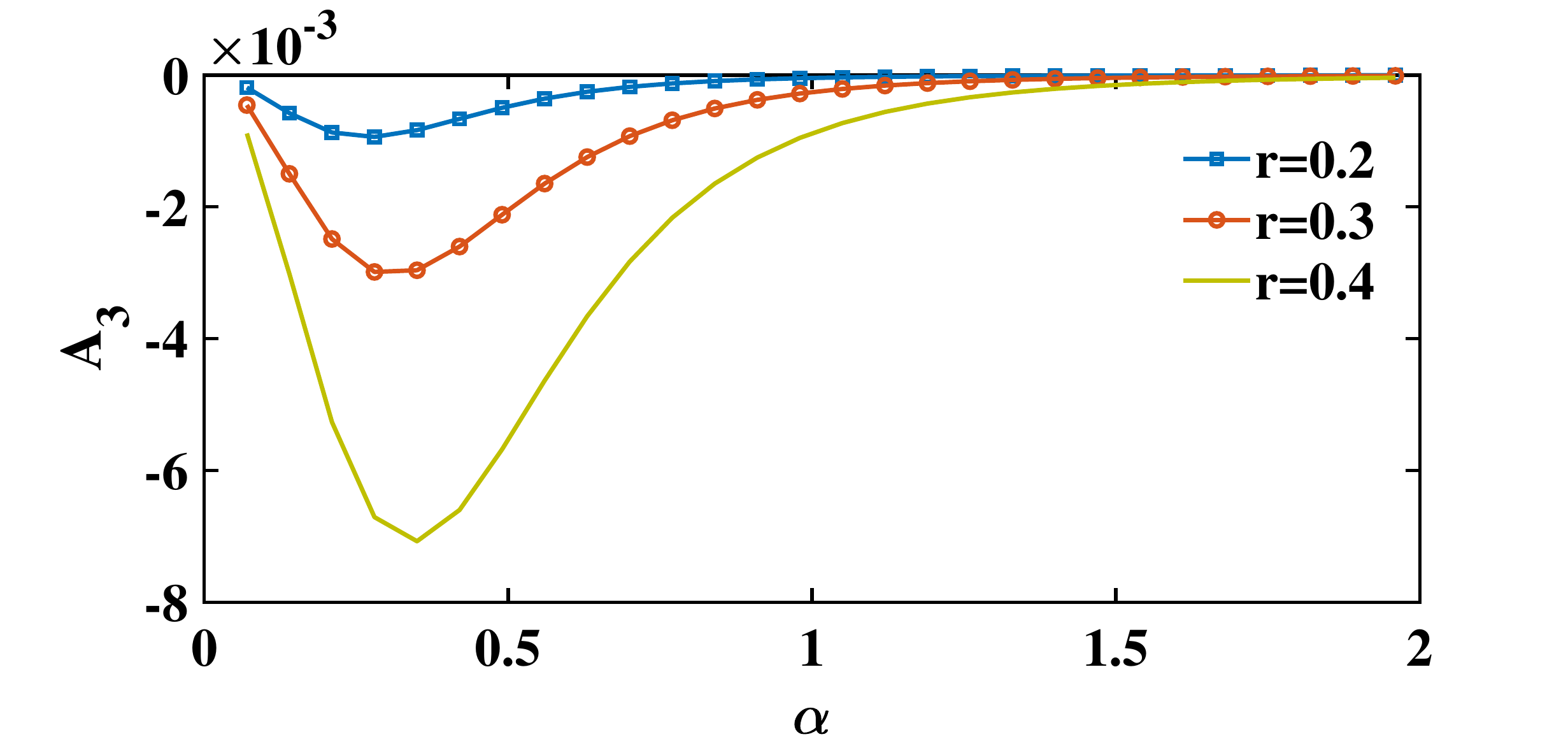}

\caption{(Color online) Variation of Agarwal-Tara parameter with respect to $\alpha$.}
\label{fig7}
\end{figure*}

The parameter $A_3$ is zero for a coherent state (classical state) and -1 for a Fock state (most nonclassical state), respectively. Thus for a nonclassical state, $A_3$ is negative and bounded by the value -1 when the state becomes maximally nonclassical. In order to investigate the nonclassicality of the superposed state in terms of $A_3$, we plot the corresponding results in Fig.~\ref{fig7}. Here $A_3$ varies between 0 to -0.008 with respect to $\alpha$ and thus depicts the presence of nonclassicality. Also, for higher values of $r$, the depth of the nonclassicality increases which is consistent with the results obtained by different moment based criteria.

\subsection{Klyshko's criterion}

Klyshko introduced a criterion to witness the nonclassicality property of a quantum state by using only three successive photon-number probabilities \cite{klyshko}. If $p_m = \braket{{m}|\rho|{m}}$ is the photon-number probability of a state having density matrix $\rho$, then the Klyshko's inequality can be written as
\begin{equation}
B(m) = (m+2)p_mp_{m+2} - (m+1)(p_{m+1})^2\,\, <\,\,0
\end{equation}
Using $p_m = N^{-1}\frac{e^{-|\alpha|^2}}{m!}{|\alpha|}^{2(m-1)}|t\,\alpha^2+r\,m|^2$, the detailed expression for $B(m)$ is
\begin{widetext}
\begin{eqnarray}
\begin{array}{rcl}
B(m) = -\frac{e^{-2|\alpha|^2}|\alpha|^{4m}}{N^2 m!(m+1)!}
r^2\Big[r^2(2m^2+4m+1)+2rt(m+1)(\alpha^2+\alpha^{\star 2})+t^2(\alpha^4+\alpha^{\star 4})\Big]
\end{array}
\end{eqnarray}
\end{widetext}

\begin{figure*}[ht]
\centering
\includegraphics[width=5.5cm]{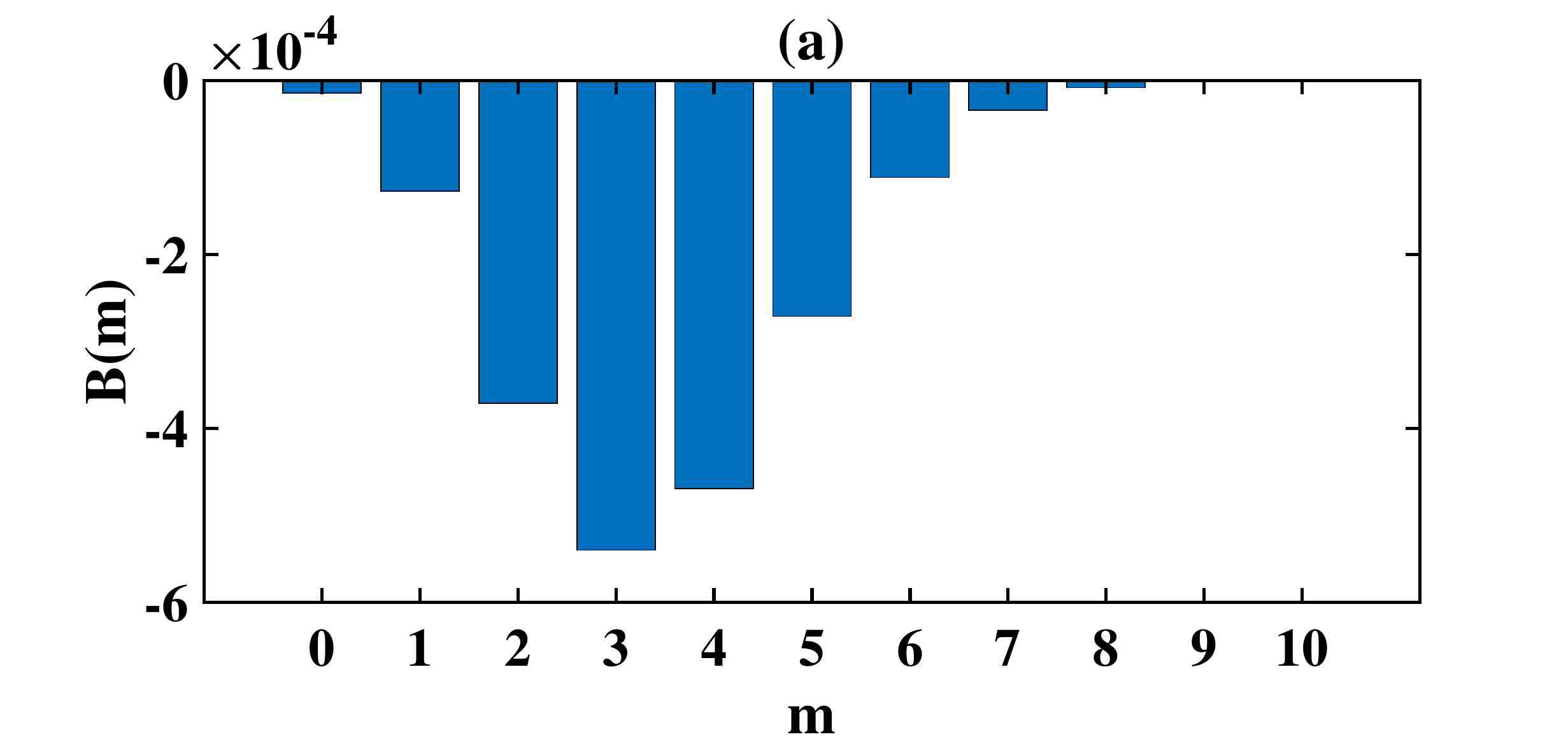}
\includegraphics[width=5.5cm]{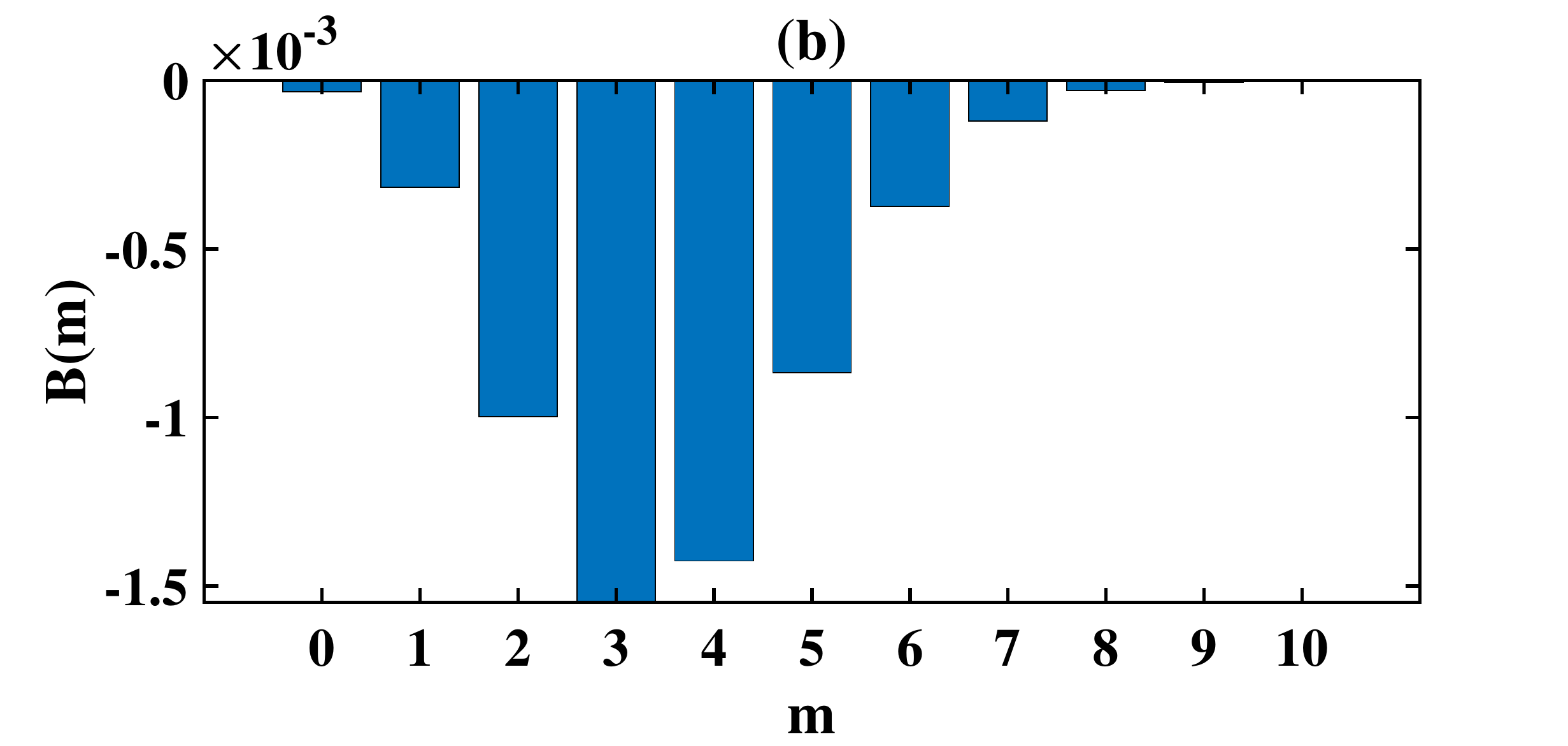}
\includegraphics[width=5.5cm]{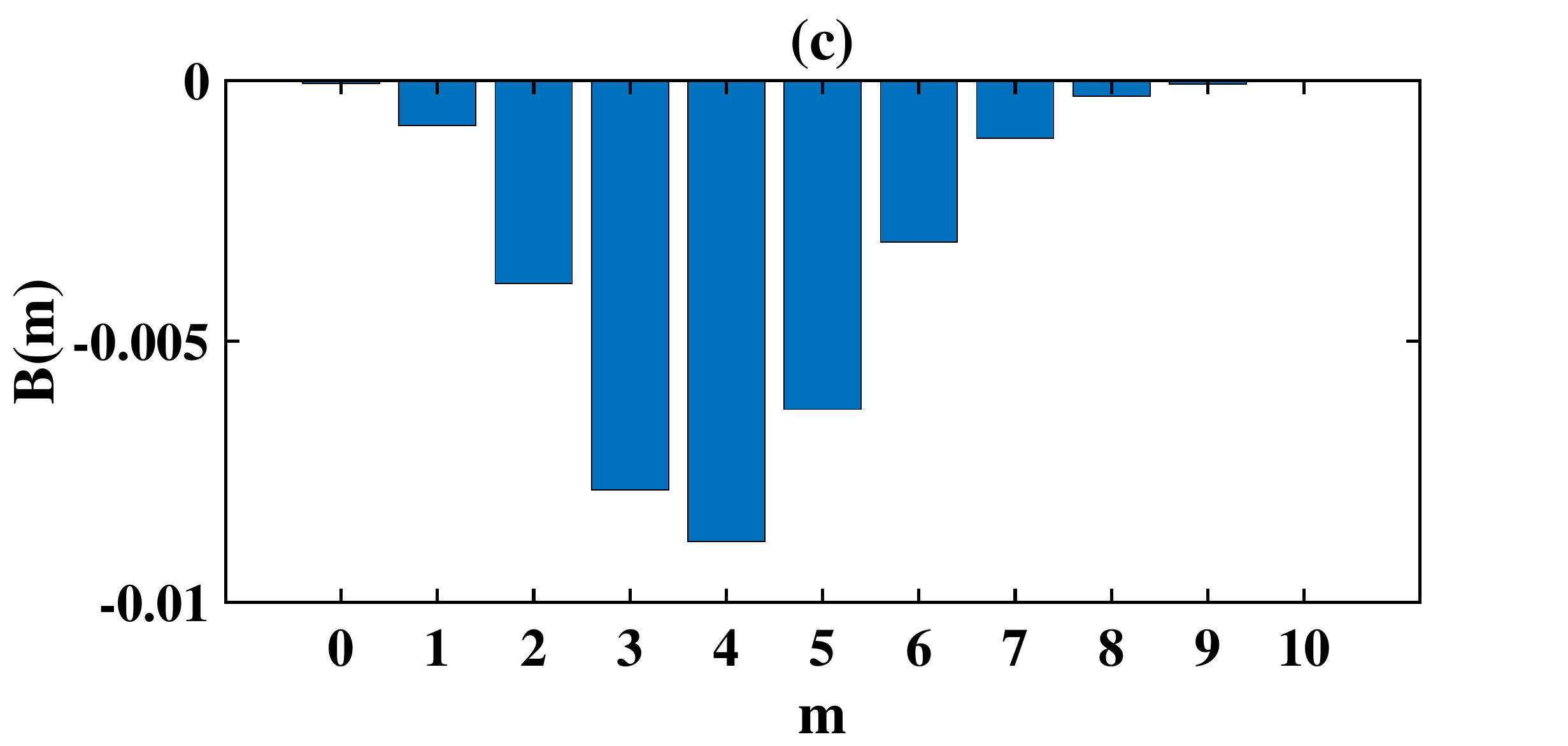}

\caption{(Color online) Illustration of Klyshko's criterion $B(m)$ as a function of $m$ with $\alpha=2$ and different values of $r$ as (a) $r=0.2$, (b) $r=0.38$ and (c) $r=0.94$, respectively.}
\label{fig8}
\end{figure*}

The advantage of the Klyshko's criterion over any other existing moment based criterion is that a very small amount of information is required. In this criterion, we need only the photon number distribution $p_n$ for the three successive values of $n$. The negative values of $B(m)$ serve as the witness of nonclassicality here. For fixed values of $\alpha$ and $r$, we observe that $B(m)$ is always negative which signifies the existence of a nonclassical photon statistics. It can be visualized that $B(m)$ becomes most negative around $m=3$.

\begin{figure*}[ht]
\centering
\includegraphics[width=5.5cm]{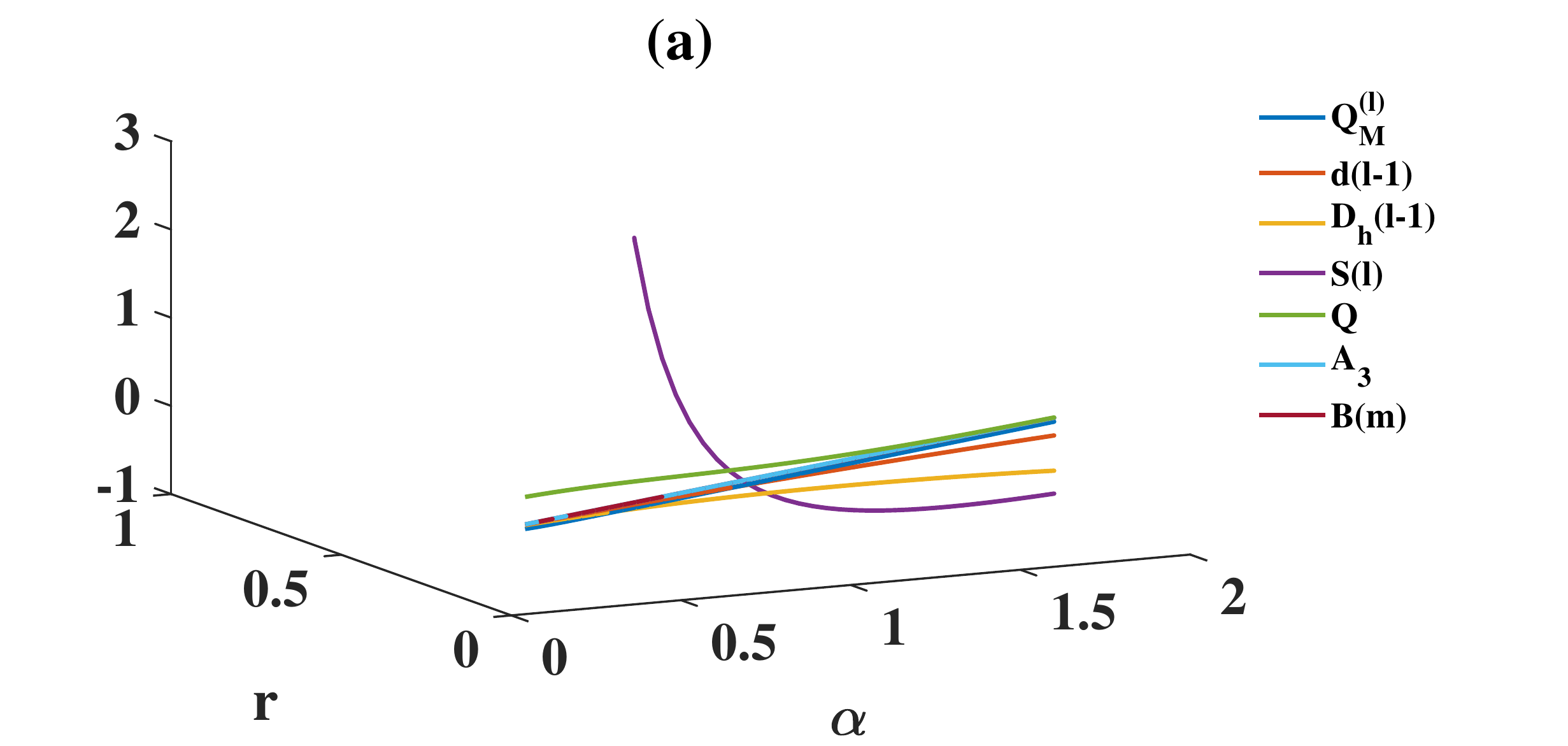}
\includegraphics[width=5.5cm]{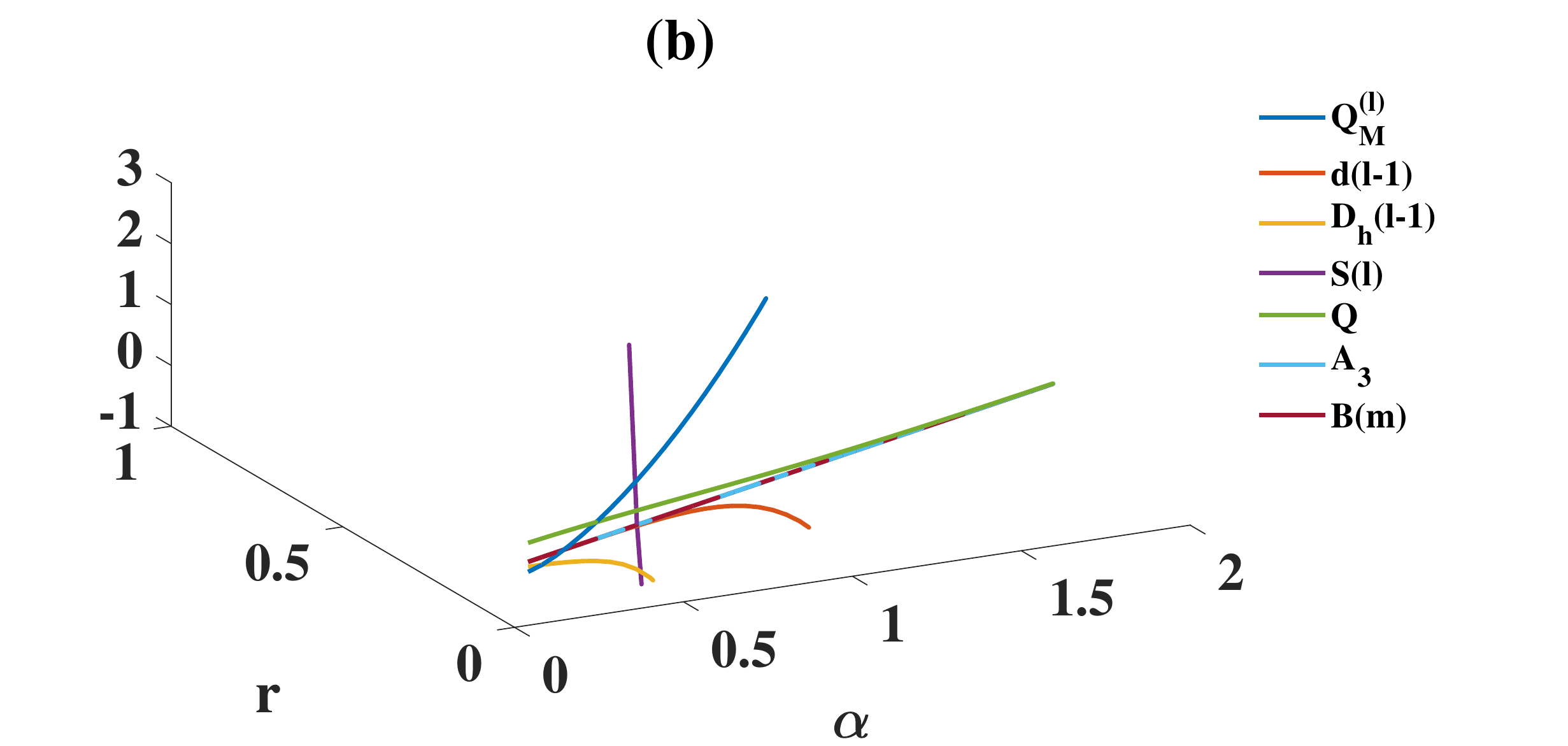}

\caption{(Color online) Domains of nonclassicality detected by seven different criteria in the $r\mbox{-}\alpha$ plane with (a) $l=2$, $\beta=0.1$ and (b) $l=4$ and $\beta=0.1$.}
\label{fig9}
\end{figure*}
Fig.~\ref{fig9} describes a comparative plot for all the seven criteria studied so far, as a function of $r$ and $\alpha$. The first figure presents the domain of nonclassicality for lower-order ($l=2$) criteria while the next one is for higher-order ($l=4$) conditions.
\section{Conclusion}
\label{sec5}

In conclusion, we have introduced a quantum state by applying a combination of two operators $a$ and $a^{\dagger}$ to a coherent state $\ket\psi$. The scalars $t$ and $r$ act as control parameters for manipulation of the nonclassical character of the output state. We have focused on the higher-order nonclassical features of the state.

In the present work, a schematic diagram is presented to realize the superposed operation $ta+r a^\dagger$. Then the quantum state $\ket\psi$ is formed by operating $ta+r a^\dagger$ over a coherent state $\ket\alpha$. A set of various measurement techniques is used here to check the existence of higher-order nonclassicality in the superposed state. It is found that higher-order Mandel's $Q_M$ parameter can identify the nonclassicality in a certain range of state parameter $\alpha$ but the lower-order cannot. The same is true for HOA. Further, it is observed that the probability of getting a bunch of photons is decreased as $r$ increases. Another higher-order nonclassicality phenomenon HOSPS is found in accordance with HOA. But in case of squeezing, the superposed state depicts the lower-order property while corresponding HOS is absent. The dependence of lower-order squeezing on phase parameter is also displayed. In addition, the nonclassical nature is also investigated through a quasiprobability $Q$ function, Agarwal-Tara $A_3$ parameter and Klyshko's criterion. All these measures (except Husimi $Q$) can detect nonclassicality. The phase-space-matrix approach, which incorporates
nonclassicality tests based on negativities of the phase-space distributions, is also applied to show the nonclassical nature of the superposed state. It is also clarified from the figures that the amount of nonclassicality increases with the control parameter $r$.

\begin{center}
\textbf{ACKNOWLEDGEMENT}
\end{center}
Deepak's work is supported by the Council of Scientific and Industrial Research (CSIR), Govt. of India (Award no. 09/1256(0006)/2019-EMR-1).

\begin{center}
\textbf{DISCLOSURES}
\end{center}

The authors declare no conflicts of interest.



\begin{thebibliography}{99}
\newcommand{\enquote}[1]{``#1''}

\bibitem{glauber} R. J. Glauber, Phys. Rev. 131 (6) (1963) 2766.
\bibitem{agar} G. S. Agarwal, Quantum Optics, Cambridge University Press, 2013.
\bibitem{klauder} J. Klauder and B. Skagerstam, Coherent States: Applications in Physics and
Mathematical Physics, World Scientific, 1985.
\bibitem{zhang} W. M. Zhang, R. Gilmore et. al., Rev. Mod. Phys. 62(4) (1990) 867.
\bibitem{bellini} A. Zavatta, V. Parigi, M. S. Kim, H. Jeong and M. Bellini M, Phys. Rev. Lett. 103 (2009) 140406.
\bibitem{xu} L. Y. Hu, X. X. Xu, Z. S. Wang and X. F. Xu, Phys. Rev. A 82 (2010) 043842.
\bibitem{chat} A. Chatterjee, Phys. Lett. A 376 (2012) 1601-1607.
\bibitem{priya1} P. Malpani, K. Thapliyal, N. Alam, A. Pathak, V. Narayanan and S. Banerjee, Ann. Phys. (Berlin) 11 (2019), 1900141.
\bibitem{pathak2} A. Pathak, Elements of Quantum Computation and Quantum Communication, Taylor and Francis, (2013).
\bibitem{tiwari} P. Tiwari, S. Dehdashti, A. K. Obeid, M. Melucci and P. Bruza, arxiv:2007.07887.
\bibitem{dajka} J. Dajka and J. Luczka, Entropy (2020) 22(2) 201.


\bibitem{tittel} W. Tittel, G. Ribordy and N. Gisin, Phys. World 11 (1998) 41-46.
\bibitem{knill} E. Knill, R. Laflamme and G. Milburn, Nature 409 (2001) 46-52.
\bibitem{carmichael} H. Nha and H. J. Carmichael, Phys. Rev. Lett. 93 (2004) 020401.
\bibitem{taka} H. Takahashi et al., Nature Photonics 4 (2010) 178.
\bibitem{yang} Y. Yang and F. L. Li, Phys. Rev. A 80 (2009) 022315.
\bibitem{kim} M. S. Kim, J. Phys. B 41 (2008) 133001.
\bibitem{tara1} G. S. Agarwal and K. Tara, Phys. Rev. A 43 (1991) 492–497.
\bibitem{filho} R. L. Filho and W. Vogel, Phys. Rev. A 54 (1996) 4560-4563.
\bibitem{manko} V. L. Ma\'{n}ko, G. Marmo, E. C. G. Sudarshan and F. Zaccaria, Phys. Scripta 55 (1997) 528-541.
\bibitem{ghosh1} A. Chatterjee and R. Ghosh, J. Opt. Soc. Am. B 33(7) (2016) 1511-1522.
\bibitem{parigi} M. S. Kim, H. Jeong, A. Zavatta, V. Parigi and M. Bellini, Phys. Rev. Lett. 101 (2008) 260401.
\bibitem{lee} S. Y. Lee and H. Nha, Phys. Rev. A 82 (2010) 053812.
\bibitem{ralph} T. C. Ralph, A. G. White, W. J. Munro and G. J. Milburn, Phys. Rev. A 65 (2001) 012314.
\bibitem{adesso} G. Adesso, F. Dell\'{A}nno, S. De Siena, F. Illuminati and L. A. M. Souza, Phys. Rev. A 79 (2009) 040305(R).



\bibitem{kiesel} T. Kiesel, W. Vogel, V. Parigi, A. Zavatta and M. Bellini, Phys. Rev. A 78(2) (2008) 021804.
\bibitem{miran} A. Miranowicz, M. Bartkowiak, X. Wang, Y. X. Liu and F. Nori, Phys. Rev. A 82(1) (2010) 013824.
\bibitem{monica} M. Bartkowiak, A. Miranowicz, X. Wang, Y. Liu, W. Leo\'{n}ski and F. Nori, Phys. Rev. A 83(5) (2011) 053814.
\bibitem{gracia} A. Pathak and M. E. Garcia, Appl. Phys. B 84(3) (2006) 479-484.
\bibitem{prakash} H. Prakash and D. K. Mishra, J. Phys. B: Atom. Mol. Phys. 39(9) (2006) 2291.
\bibitem{hillery} M. Hillery, Phys. Rev. A 36(8) (1987) 3796.
\bibitem{hong1} C. K. Hong and L. Mandel, Phys. Rev. Lett. 54(4) (1985) 323.
\bibitem{allevi1} A. Allevi, S. Olivares and M. Bondani, Int. J. Quant. Inform.  10(08) (2012) 1241003.
\bibitem{jack} J. Pe\v{r}ina Jr, V. Mich\'{a}lek and O. Haderka, Phys. Rev. A 96(3) (2017) 033852.
\bibitem{alam2} N. Alam, K. Thapliyal, A. Pathak, B. Sen, A. Verma and S. Mandal, arXiv:1708.03967 (2017).
\bibitem{kishore1} K. Thapliyal, A. Pathak, B. Sen and J. Pe\v{r}ina, Phys. Rev. A 90(1) (2014) 013808.
\bibitem{kishore4} K. Thapliyal, A. Pathak, B. Sen and J. Pe\v{r}ina, arXiv:1710.04456 (2017).
\bibitem{alam1} N. Alam, N., A. Verma and A. Pathak, Phys. Lett. A 382 (2018) 1842–1851.
\bibitem{kishore2} K. Thapliyal, N. L. Samantray, J. Banerji and A. Pathak, Phys. Lett. A 381(37) (2017) 3178–3187.
\bibitem{alam3} N. Alam and S. Mandal, Opt. Commun. 359 (2016) 221-233.
\bibitem{martin} M. Bohmann, E. Agudelo and J. Sperling, Quantum 4 (2020) 343.

\bibitem{himadri1} A. Chatterjee, H. S. Dhar and R. Ghosh, J. Phys. B: At. Mol. Opt. Phys. 45 (2012) 205501.
\bibitem{wenger} J. Wenger, R. Tualle-Brouri and P. Grangier, Phys. Rev. Lett. 92 (2004) 153601.
\bibitem{zavatta} A. Zavatta, S. Viciani and M. Bellini, Science 306 (2004) 660.



\bibitem{sudarshan} E.C.G. Sudarshan, Phys. Rev. Lett. 10 (7) (1963) 277.
\bibitem{wigner} E. P. Wigner, Phys. Rev. 40(5) (1932) 749.
\bibitem{kenfack} A. Kenfack and K. \.{Z}yczkowski, J. Opt. B: Quantum Semiclass. Opt. 6(10) (2004) 396.
\bibitem{husimi} K\^{o}di Husimi, Proc. Phys. -Math. Soc. Japan. 3rd Ser. 22(4) (1940) 264–314.
\bibitem{stephen} N. L\"{u}tkenhaus and S. M. Barnett, Phys. Rev. A 51(4) (1995) 3340.



\bibitem{allevi} A. Allevi, S. Olivares and M. Bondani, Phys. Rev. A 85(6) (2012) 063835.\\
A. Verma, N. K. Sharma and A. Pathak, Phys. Lett. A 372(34) (2008) 5542–5551.
\bibitem{mandel} L. Mandel, Optics Letters  4(7) (1979) 205–207.
\bibitem{sanjib} S. Dey and V. Hussin, Phys. Rev. A 93 (2016) 053824.
\bibitem{moya1} J. M. Vargas Mart\'{i}nez, H. Moya-Cessa and M. F. Guasti, Rev. Mexi. De F\'{i}sica E 52(1) (2006) 13-16.
\bibitem{stegun} M. Abramowitz and I.A. Stegun, Handbook of Mathematical functions (Dover, New York, 1968)
\bibitem{ching} C. T. Lee, Phys. Rev. A 41(3) (1990) 1721.



\bibitem{amit} A. Verma and A. Pathak, Phys. Lett. A 374(8) (2010) 1009–1020.
\bibitem{hong} C. K. Hong and L. Mandel, Phys. Rev. A 32(2) (1985) 974.

\bibitem{kishore3} K. Thapliyal, S. Banerjee, A. Pathak, S. Omkar and V. Ravishankar, Annal. Phys. 362 (2015) 261-286.
\bibitem{slus} S. Slussarenko and G. J. Pryde, Appl. Phys. Rev. 6 (2019) 041303.
\bibitem{yadin} B. Yadin, F. C. Binder, J. Thompson, V. Narasimhachar, M. Gu and M. S. Kim, Phys. Rev. X 8 (2018) 041038.
\bibitem{kwon} H. Kwon, K. C. Tan, T. Volkoff and H. Jeong, Phys. Rev. Lett. 122 (2019) 040503.
\bibitem{shah1} F. Shahandeh, A. P. Lund and T. C. Ralph, Phys. Rev. Lett. 119 (2017) 120502.
\bibitem{shah2} F. Shahandeh, A. P. Lund and T. C. Ralph, Phys. Rev. A 99 (2019) 052303.



\bibitem{girish} G. S. Agarwal, K. Tara, Phys. Rev. A 46(1) (1992) 485.
\bibitem{priya} P. Malpani, K. Thapliyal, N. Alam, A. Pathak, V. Narayanan and S. Banerjee, Opt. Commun. 459 (2020) 124964.
\bibitem{klyshko} D. N. Klyshko, Phys. Lett. A 213(1-2) (1996) 7-15.
\end{thebibliography}
\end{document}